# Gradient Extension of Classical Material Models
## From Nuclear & Condensed Matter Scales to Earth & Cosmological Scales

**Elias C. Aifantis**[*]

School of Engineering, Aristotle University of Thessaloniki, Thessaloniki GR-54124, Greece[†]

**\* Correspondence:**
Corresponding Author
mom@mom.gen.auth.gr

**Abstract**

The various mathematical models developed in the past to interpret the behavior of natural and manmade materials were based on observations and experiments made at that time. Classical laws (such as Newton's for gravity, Hooke's for elasticity, Navier-Stokes for fluidity, Fick's/Fourier's for diffusion/heat transfer, Coulomb's for electricity, as well as Maxwell's for electromagnetism and Einstein's for relativity) formed the basis of current technology and shaping of our civilization. The discovery of new phenomena with the aid of recently developed experimental probes have led to various modifications of these laws across disciplines and the scale spectrum: from subatomic and elementary particle physics to cosmology and from atomistic and nano/micro to macro/giga scales. The emergence of nanotechnology and the further advancement of space technology are ultimately connected with the design of novel tools for observation and measurements, as well as the development of new methods and approaches for quantification and understanding. The paper first reviews the author's previously developed weakly nonlocal or gradient models for elasticity, diffusion and plasticity within a unifying internal length gradient (ILG) framework. It then proposes a similar extension for fluids and Maxwell's equations of electromagnetism. Finally, it ventures a gradient modification of Newton's law of gravity and examines its implications to some problems of elementary particle physics, also relevant to cosmology. Along similar lines, it suggests an analogous extension of London's quantum mechanical potential to include both an "attractive" and a "repulsive" branch. It concludes with some comments on a fractional generalization of the ILG framework.

## 1    Introduction

In a recent chapter in Advances of Applied Mechanics [1], a detailed account is presented of the author's internal length gradient (ILG) mechanics framework. It is based on the assignment of internal lengths/ILs (associated with the local geometry/topology of material nano/microstructures) as scalar multipliers of extra Laplacian terms that are introduced to account for heterogeneity effects and weak nonlocality. Related background work for this framework can be found in the references quoted therein, as well as in earlier published articles by the author and his coworkers [2-10].

---

[†] Emeritus Professor. Also Emeritus Professor of Michigan Technological University, Houghton, MI 49931, USA & Mercator Fellow Friedrich-Alexander University, Erlangen-Nüremberg, Fürth 90762, Germany.

The motivation for the development of the author's initial continuum mechanics-based ILG framework was the need for describing deformation pattern-forming instabilities that emerge when an externally applied stress reaches a certain threshold. Beyond that threshold, the evolution equations governing the system's homogeneous response were becoming ill-posed and further analysis was not possible. The method proposed earlier by the author to overcome the difficulty for macroscopic deformation and fracture instabilities, was to introduce higher-order gradients (in the form of Laplacians) in the constitutive equations and corresponding ILs accounting for the heterogeneity of the underlying micro/nano structures. The resulting differential equations eliminate ill-posedness, estimate the width/spacings of deformation bands, dispense with the mesh-size dependence in finite element calculations, and remove stress/strain singularities at crack tips. A similar approach has been employed by the author for higher-order diffusion and heat conduction theories, as well as for phase transitions by revisiting Van der Waals theory of liquid-vapor interfaces and Cahn-Hilliard theory of spinodal decomposition [11] through the introduction of chemical ILs. In these works, mechanical and chemical ILs were treated separately as phenomenological parameters, depending on the material local configuration and scale of observation. Their calibration and/or estimation was left to numerical and/or laboratory experiments. Moreover, statistical features emerging at sub-macroscopic scales were not considered. A preliminary effort to address these issues has been outlined in [1], and further been elaborated upon herein. In particular, the powerful multiscale technique of Kevrekidis [12] – the equation-free method/EFM – can be utilized for the hierarchical calibration of mechanical ILs. Their experimental estimation, usually inferred from "indirect" measurements of spatio-temporal features (width/spacing/velocity of deformation bands) and related size effects, can be based on "direct" measurements through novel nanoindentation/NI tests by monitoring the local strain gradients at various indentation depths.

The enhancement of the above deterministic ILG considerations to include stochastic effects associated with internal stress fluctuations that manifest as stress drops/strain bursts in micro/nanopillar experiments and popins/popouts in nanoindentation tests, may be pursued along the lines also outlined by in [1]. Corresponding gradient-stochastic models can be derived to capture incipient plasticity and serrated stress-strain graphs, as well as to determine statistical features such as fractal dimensions/FDs and probability density functions/PDFs. This task can be carried out by employing Tsallis q-statistics [13] – based on "nonextensive entropy" (as opposed to Boltzmann-Gibbs/B-G "extensive thermodynamics") – resulting to q-dependent multifractal spectra and q-dependent PDFs, as well as q-generalization of B-G universal power laws. Novel NI tests may be conducted on multiple specimen sites and at different penetration depths for the determination of q-distributions by recording the observed popins/popouts and comparing them with corresponding determinations from micro/nanopillar serrated stress-strain curves in other Labs. Such a statistical mechanics enhanced ILG framework may also be conveniently employed to consider the Portevin Le Chatelier/PLC and Persistent slip band/PSB plastic instabilities, along with related size effects, as outlined below.

At very small scales, mechanical and chemical effects are often equipresent, and an extended chemomechanical ILG framework is necessary in order to consider higher-order IL couplings, as suggested in [1]. In view of the fact that mechanical and chemical ILs are introduced as scalar multipliers of corresponding Laplacian terms, it turns out that such coupled chemomechanical formulation is appealing and robust. Since in mathematical biology models cells are represented by scalar concentration fields (i.e. in the same way as chemical species), the formulation could be easily adapted for the description of higher-order couplings between mechanical and biochemical ILs. Such an extended ILG mechanics framework, including synergistic effects between mechanical and chemical or biological ILs, can be employed to consider chemomechanical instabilities in LiB anodes and biomechanical instabilities in brain tumors, as also outlined below.

As mentioned above, we conclude this introductory section by summarizing main results of the ILG framework and its potential to be employed for considering a variety of problems of current or emerging interest as follows:

- *Plastic Instabilities and Size Effects*: Recent experiments at micro/nano scales [14] have revealed a strong dependence on specimen size. Ongoing work in several Labs has revealed, in particular, that PLC and PSB instabilities may be suppressed when the ratio of the specimen size over the internal length is reduced below a certain threshold. The previous deterministic ILG models earlier advanced by the author and collaborators for these instabilities at macro/meso scales can be revisited and evaluated for "small-volumes" and strain localization phenomena observed in NC and UFG polycrystals. New combined gradient-stochastic ILG models for both PLC and PSB instabilities can be employed to capture spatio-temporal periodicity, fractality, and transition to chaos. FDs for the observed deformation bands and PDFs for the recorded serrations in stress-strain curves can be determined through Tsallis q-statistics.

- *Chemomechanical Instabilities & Lithiation Fronts in LiB Anodes:* A deterministic version of our coupled chemomechanical ILG framework can be employed to address chemostress damage instabilities in nanostructured LiB anodes leading to cracking and capacity fade during Li insertion/de-insertion under electrochemical cycling. This is due to the huge local volume expansions (up to 400%) and associated internal stress generation occurring in Si active particles during lithiation [15]. A related issue is to understand the size/stress dependence of lithiation reaction, as well as the propagation of stress-assisted lithiated fronts which controls battery efficiency. The interplay between higher-order mechanical and chemical ILs has not yet been sufficiently considered to address these chemomechanical instabilities in LiBs, despite of their wide use in microelectronics, laptops and electric car technologies.

- *Biomechanical Instabilities & Cancer Growth/Metastasis in Human Brain:* A striking analogy exists between the Walgraef-Aifantis/W-A model [16] of dislocation patterning in PSBs and the Go or Grow/GoG model for glioblastoma cancer cells [17]. Both processes are described by similar reaction-diffusion/R-D type equations for mobile-immobile dislocations in the W-A model (under the action of applied stress) and the motile-immotile cancer cells in the GoG model (under the action of internally generated stress). Such internal stress effects have not been explicitly accounted for in the GoG model, despite of the fact that Murray [18] – the father of modern mathematical biology – had already introduced cell-tractions and corresponding strain gradients (in the form of Laplacians, as in the author's work; see, for example, the related discussions in [1]) to revisit Turing's seminal R-D work of morphogenesis. The interplay between higher-order mechanical and biochemical ILs in the GoG model can be studied, and the role of internal stress can thus be evaluated. The results can provide new insight on brain cancer progression and potential therapeutic procedures.

## 2    State-of-the Art: Previous Literature & Current State of Affairs

An extensive bibliography on gradient theories has already been mentioned that can be found in [1]. Specific aspects pertaining to the present review and related work on continuum mechanics models at micro/meso/macro scales are discussed in this section. For the convenience of the reader we present this section in two parts. In the first part we provide background on relatively recent phenomenological strain gradient models that have been developed to capture mechanically-induced plastic instabilities and size effects under the action of applied loads. In the second part we provide a brief account of earlier, more fundamental work on phase transitions which was a direct motivation for the author's initial ILG deterministic models, as well as for the development of coupled chemomechanical/ biomechanical ILG models to be used for addressing instabilities in LiBs and brain tumors.

## 2.1   Plastic Instabilities & Size Effects

The term "material instabilities" and "dislocation patterning" were quoted by the author in the mid 1980s [2-5,19,20] to denote the self-organization of localized strain bands and dislocations in deforming solids. Various gradient dislocation dynamics and gradient plasticity models were generated to deal with dislocation pattern formation and shear band thickness/spacing evolution, as well as for interpreting size effects [21-22]. Soon afterwards, in the beginning of the 1990s [2-5,19,23], the author incorporated the Laplacian of the Hookean stress into the standard constitutive equation of linear elasticity, to remove the singularities from dislocation lines and crack tips. Some of this work is reviewed in a number of book chapters [14] by leading authors in the field.

Subsequently, or in parallel to the above developments, other types of gradient models have been developed such as the Fleck-Hutchinson and the Gao-Nix-Huang strain gradient theories, as well as improved gradient theories taking into account surface effects (Gudmundson, K.E. Aifantis/Willis, Polizzotto, Voyiadjis/Bammann et al). As an alternative to the initial W-A phenomenological model for dislocation patterning, a substantial effort has been initiated based on discrete dislocation dynamics (DDD) modeling (Kubin/Ghoniem/Bulatov and coworkers). Due to computational limitations of DDD for obtaining dislocation patterns and motivated by the initial W-A model of [16], alternative dislocation density based methods or continuum dislocation dynamics (CDD) have been pursued (Groma/El Azab/Zaiser and coworkers). Related references connected to the above named authors can be found in the bibliography listed in [21] and [24]. In this connection, it is pointed out that our gradient elasticity model has recently been successfully utilized by N. Ghoniem's group in UCLA to dispense with near-core singularities causing code-malfunctioning in 3D discrete dislocation dynamics simulations [25]. Moreover, our related non-singular strain/stress crack tip solutions have been successfully used by P. Isaksson's group in Uppsala to interpret experimental measurements on crack-tip profiles in micro-heterogeneous materials such as solid foams and bone tissue [26].

With the exception of the author's preliminary efforts described in [1], all the above works on gradient models for addressing plastic instabilities and size effects do not account for internal stress/structural defect fluctuations and synergistic gradient-stochastic effects. There are no attempts for a hierarchical IL calibration through EFM multiscale simulations and novel NI tests. The same holds for the use of Tsallis q-statistics to determine FDs and PDFs. All these are open inter-dependent issues that need to be addressed. It is also pointed out that none of the above gradient elasticity/plasticity models incorporate diffusion and chemical reaction effects. An exception can again be found in [1] where higher-order chemomechanical IL couplings are discussed. This issue needs further addressed to consider chemomechanical and biomechanical instabilities as described below.

## 2.2 Chemomechanical Instabilities in LiBs & Biomechanical Instabilities in Brain

The main reason that the author's Laplacian-based ILG models can be easily extended to include chemical and biochemical ILs is due to the fact that their motivation stems from his earlier treatment with Serrin [11$^{a,b}$] on Van der Waals theory of fluid interfaces, which was also the predecessor of Ginzburg-Landau theory of phase transitions and Cahn-Hilliard theory of spinodal decomposition [11]. This is in contrast to the aforementioned Fleck-Hutchinson and related strain gradient models which were motivated by Cosserat-type generalized continuum mechanics theories that do not contain explicitly the Laplacian and, thus, they do not exhibit the corresponding mathematical and physical properties that this operator implies. On the other hand, chemical reactions and phase transformations have traditionally been treated with R-D equations involving the Laplace operator. The fact that both mechanical and chemical or biochemical ILs can be treated on the same footing through the introduction of the corresponding Laplacians allows for a robust formulation of a chemomechanical and biomechanical ILG framework that can be used to consider corresponding instability phenomena in LiB anodes and brain glioblastomas, respectively.

There is a large number of recent articles on LiB capacity fade due to colossal volume changes in anodes (up to 400% for Li-Si based anodes) during lithiation/delithiation [15]. While in some of these works diffusion and coupled deformation-diffusion effects have been accounted for, higher-order strain gradients and corresponding mechanical ILs have not been considered. An exception is the recent article by the author and coworkers in [19] employing strain gradients and mechanical ILs to model size effects in LiB anodes, as well as in [20] employing both mechanical and chemical ILs to model the propagation of lithation fronts. This work can be used as a guide to develop criteria for the most optimum nanocomposite configuration (size/spacing of active Si-nanoparticles) for LiB anodes to prevent cracking and/or accelerate lithiation/delithiation.

Similarly to the case of LiBs, there is an abundance of mathematical models for brain cancer. However, related ILG models accounting for internal stress gradient effects due to tumor growth and cancer cell migration/proliferation are missing. This is also true for the aforementioned GoG [10] phenotypic plasticity model of cancer cell migration and its impact on tumor progression. It was found that low-grade tumor micro-ecology potentially exhibits an emergent Allee effect, i.e. a critical tumor cell density implying tumor growth or control. The precise quantification of this critical tumor cell density could be a relevant prognostic criterion for the tumor fate through biopsy measurements. It was shown that the GoG mechanism explains the fast tumor recurrence time of high-grade brain tumors after resection. These findings can be re-evaluated by transferring the stability analysis results earlier derived for the W-A model (of mobile-immobile dislocations) to the GoG model (of motile-immotile cells), by also considering stochastic heterogeneity and internal stress gradient effects.

## 3 Mechanics ILG Formulation through Continuum & Statistical Mechanics

*ILs in Elasticity/Plasticity & Diffusion/Gradient Dislocation Dynamics:* For elastic deformations, the term $\ell_\varepsilon^2 \nabla^2 [\lambda \varepsilon_{mm} \delta_{ij} + 2G\varepsilon_{ij}]$ – where $\ell_\varepsilon$ denotes elastic IL, $\varepsilon_{ij}$ is the elastic strain ($\varepsilon_{ij} = \frac{1}{2}[u_{i,j} + u_{j,i}]$; $u_i$ designates displacement), and $(\lambda, G)$ are the Lamé constants – is incorporated into classical Hooke's law. Previous results of the author and his coworkers (see [1] and references quoted therein) show that the resulting ILG model can eliminate stress/strain singularities from dislocation/disclination lines and crack tips, as well as interpret elastic size effects. Similarly, the term $\ell_p^2 \nabla^2 \gamma^p$ – where $\ell_p$ denotes plastic IL and $\gamma^p = \int \dot\gamma^p dt$ ($\dot\gamma^p = \sqrt{2\dot\varepsilon_{ij}^p \dot\varepsilon_{ij}^p}$) is the equivalent plastic strain with $\varepsilon_{ij}^p$ denoting the plastic strain tensor – is introduced in the classical von-Mises yield condition or the flow rule to derive differential equations that remain well-posed into the unstable flow regime. Previous results of the author and his coworkers (e.g. [1] and refs therein) show that the resulting ILG model can determine shear band widths and spacings, as well as interpret plasticity induced size effects in micro-torsion/bending and micro/nano indentation experiments. For elastic deformations at the atomic scale (near dislocation lines in crystals), $\ell_\varepsilon$ relates to the subatomic configuration and electronic state (through DFT calculations), while at the microscale $\ell_\varepsilon$ relates to particle size/spacing (through MD simulations). For plastic deformations at micro/meso scales (deformation bands, dislocation cells), $\ell_p$ relates to dislocation source distance/pileup length/grain size (through DDD simulations). This suggests that our earlier practice of treating the ILs as "fitting" constants needs to be revised and consider them as evolving parameters in the course of deformation. This point of view can be adopted for exploring the IL-dependence on the current state of deformation and underlying micro/nanostructural configuration, also in relation to the size of the volume considered.

For diffusion problems, the ILs enter through the additional term $\ell_d^2 \nabla^2 j_i$ which generalizes the classical Fick's law ($\ell_d$ is a diffusional internal length and $j_i$ denotes the diffusion flux) in a manner similar to the Cahn-Hilliard theory [11] for spinodal decomposition. For collective dislocation phenomena, the IL enters through the extra Laplacian term $D\nabla^2 \rho$, where $\rho$ denotes dislocation density and $D$ is an "effective" diffusion-like transport coefficient. Unlike random diffusion processes however, $D$ is a strain rate driven parameter. Since the strain rate depends (through Orowan's equation) linearly on the average dislocation velocity, which, in turn relates to the local stress, the coefficient $D$ is treated as a stress-dependent parameter that relates to individual dislocation interactions. It is noted, in this connection, that the pioneering W-A model for dislocation patterning which is based on such type of $D\nabla^2 \rho$ terms for the mobile and immobile dislocation densities was initially criticized for the phenomenological origin of these Laplacian terms. However, recent work on continuum dislocation density based dynamics (CDD) – in contrast to discrete dislocation based dynamics (DDD) simulations which were unable to produce dislocation patterns – generate such type of Laplacian terms which are necessary for dislocation pattern formation interpretations.

*Stochasticity and Tsallis q-Statistics:* The enhancement of the above discussed ILG deterministic models through the incorporation of stochastic terms is necessary in order to account for the heterogeneity and fluctuations of internal stresses, as well as deformation-induced random micro/nanostructures. The resulting combined gradient-stochastic models can capture the observed behavior at micro/nano scales, including size dependent serrated stress-strain graphs and intermittent plasticity phenomena. Some initial results along this direction have recently been reported by the author and his coworkers [30] by resorting to empirical Weibull distribution functions, as also reviewed in [1]. This approach can be adopted to describe existing experimental data on stress drops/strain jumps routinely observed in micro tension/compression and nanoindentation laboratory tests. An additional issue that can be explored here is to employ time-dependent probability distributions guided by our earlier [31] and most recent [32] work based on the formalism of stochastic differential equations.

A convenient way to consider the competition between deterministic gradient and random effects is to introduce (in analogy to Wiener processes in statistical mechanics) an additive stochastic term of the form $h(\gamma) g(x)$; $\langle g(x) g(x') \rangle = l_{corr} \delta(x-x')$ – with $l_{corr}$ denoting a correlation length, and $\delta$ being the usual Dirac delta function – into the gradient expression of the flow stress. This is not an arbitrary assumption but emerges generically if one aims at a description above the scale of the discrete substructure which defines the correlation length – i.e. within a continuum model. The delta function then simply emerges because the individual volume elements of the continuum theory are effectively uncorrelated. The function $h(\gamma)$ also covers the limiting case where only the material parameters fluctuate while the evolution is deterministic (e.g. in the case of flow stress fluctuations due to fluctuating grain orientation or in the presence of a chemical environment).

Standard deterministic ILG models cannot provide any information on measured statistical aspects of plastic deformation, such as fractal dimensions for deformation patterns; power-law exponents for dislocation avalanches [33]; and strain bursts recorded during nanoindentation [34] or micro/nanopillar compression tests [14,30]. When differential equations cannot be invented to interpret experimental data and simulations, system characterization is left to statistical analyses for establishing fractality and universal power-laws. In many cases, however, the usual power-laws based on Boltzmann-Gibbs statistics exclude the regime of low intensity-high probability events. Tsallis q-statistics [13] based on nonextensive entropy thermodynamics remove this difficulty and can be employed here to analyze intermittent plasticity and deformation patterned images obtained experimentally. This information can also allow the construction of appropriate PDFs to be used in the aforementioned combined gradient-stochastic models. Tsallis nonextensive (non-additive) q-entropy reads $S_q = k\left(1 - \sum_i p_i^q\right)/(q-1)$ and by letting $q \to 1$ recovers the familiar Boltzmann-Gibbs extensive entropy. Corresponding q-distribution functions (q-Gaussian, q-exponential, q-Weibull) functions are obtained, which for $q \to 1$ reduce to their standard counterparts.

## 4    ILG Applications: Mechanics, ChemoMechanics, and BioChemoMechanics

In this section we discuss applications of the ILG framework to describe deformation instabilities and intermittent plasticity phenomena, as well as chemomechanical instabilities in lithium-ion battery anodes and tumor glioblastomas

### 4.1    Mechanical Deformation Instabilities & Intermittent Plasticity

In this subsection we briefly discuss earlier developed ILG deformation models that were used to capture two types of propagating and stationary instabilities in metallic specimens under monotonic or cyclic applied loads. As the specimen size decreases these instabilities may be suppressed or manifest in a more complex non-deterministic manner when stochastic effects appear on equal footing as deterministic ones. This is the case for micro/nano pillar deformation where intermittent plasticity prevails and combined gradient-stochastic models are needed for interpreting size-dependent serrated stress-strain curves.

In order to provide insight on the applicability of ILG framework to capture propagating plastic deformation bands routinely observed in Al-Mg alloy specimens under tension, we list below an initial strain gradient model equation used by the author and coworkers for that purpose. It reads

$$\sigma = h\varepsilon + f(\dot{\varepsilon}) + c\varepsilon_{xx}, \tag{1}$$

where $\sigma$ denotes stress, $\varepsilon$ denotes strain, $h$ a hardening modulus, $f(\dot{\varepsilon})$ a non-monotone function with a branch of negative slope modeling strain rate softening, and the gradient coefficient $c$ (units $[\text{m}]^2 \times [\text{sec}]$) is a phenomenological constant. For constant stress rate tests ($\sigma = \dot{\sigma}_0 = h\dot{\varepsilon}_s$) and travelling wave solutions $\dot{\varepsilon} = z(x - Vt)$ – where $x$ denotes the spatial coordinate, $t$ time, and $V$ the band propagation velocity – we obtain the following Lienard type nonlinear equation

$$Z_{\eta\eta} - \mu f'(Z) Z_{\eta} + (Z - Z_s) = 0, \qquad (2)$$

where $\eta = \sqrt{h/c}(x - Vt)$, $\mu = V/\sqrt{ch}$. This equation exhibits periodic solutions for propagating strain rate bands traveling through the specimen with constant velocity. It also leads to the staircase stress-strain graphs [3]. This model, which may be considered as a predecessor of later developed more elaborate PLC models, can be revisited for a strain-dependent gradient coefficient $c$ to account for the observed increase of the strain jumps in the course of deformation. In addition, it can be used for applied constant strain-rate conditions to interpret serrated stress-strain curves exhibiting stress drops (instead of strain jumps). Internal fluctuations can be accounted for by introducing a stochastic term in Eq. (1) for the constitutive expression of the gradient stress. The resulting combined gradient-stochastic model can be evaluated according to the method discussed below to interpret non-deterministic serrations and intermittent plasticity phenomena in micro/nanopillar tests. Statistical characteristics for the serrations and corresponding PDFs can be obtained through Tsallis nonextensive q-entropy procedures. Additional typical experimental results for PLC bands and serrations in NC and UFG polycrystals can be analyzed in a similar way, as in the recent work of the author and coworkers [35].

*Stationary Persistent Slip Bands/PSBs:* Next, we briefly discuss the model equations describing the periodic ladder structure of stationary PSBs. The initial W-A model for the densities of immobile $(\rho_i)$ and mobile $(\rho_m)$ dislocations reads

$$\dot{\rho}_i = g(\rho_i) + D_i \nabla^2_{xx} \rho_i - h(\rho_i, \rho_m) \quad ; \quad \dot{\rho}_m = D_m \nabla^2_{xx} \rho_m + h(\rho_i, \rho_m), \qquad (3)$$

where $(D_i, D_m)$ denote transport stress-dependent gradient coefficients for the two dislocation populations; $h(\rho_i, \rho_m)$ is an exchange term modeling dislocation reactions of the form $h(\rho_i, \rho_m) = \beta \rho_i - \gamma \rho_m \rho_i^2$; and $g(\rho_i)$ is a generation term for immobile dislocations. The coefficients $(\beta, \gamma)$ depend on stress with $\beta$ measuring the rate of production of mobile dislocations at the expense of immobile, and $\gamma$ measuring the rate of immobilization of mobile dislocations by immobile dipoles. Since the stress remains constant during PSB formation, all these model coefficients may be assumed as constants. Then, linear stability analysis of Eqs (3) around an equilibrium homogeneous state $(\rho_i^0, \rho_m^0)$ results to a Turing instability for a critical value of the bifurcation parameter $\beta = \beta_c = \left[ \sqrt{\alpha} + \rho_i^0 \sqrt{\gamma(D_i/D_m)} \right]^2$, where $\alpha = -g'(\rho_i^0)$. The critical wave number $q_c$ is given by the expression $q_c = \left[ \alpha \gamma \rho_i^2 / D_i D_m \right]^{1/4}$ and the corresponding critical wavelength $\lambda_c = 2\pi/q_c$ turns out to be of the same order of magnitude as in the experiments.

The above linear stability results were obtained for infinite domains, i.e. for specimen sizes much larger than the internal length. For finite size specimens, corresponding linear and nonlinear stability results were obtained recently by the author and coworkers [16c, 29]. The periodic ladder structure of PSBs is revealed again but it now turns out that below a critical specimen size comparable to the internal length, the PSB instability is suppressed. This size-dependence is consistent with recently obtained experimental results [36]. Stochasticity can also be introduced in Eq. (3) and the implication of a corresponding gradient-stochastic W-A model can be readily investigated.

An additional issue that can be considered is the coupling of the W-A model with an equation for the local stress $\tau$ related to the macroscopic stress $\sigma$ through the gradient expression

$$\tau - l_\tau^2 \nabla^2 \tau = \sigma, \tag{4}$$

where now the model parameters in Eq. (3) depend on $\tau$ rather than $\sigma$. This could offer an alternative simpler way to arrive at the result of the plateau stress during PSB formation than the method followed earlier by the author and coworkers [31].

*Intermittent Plasticity:* We conclude this section with some preliminary results on intermittent plastic instabilities by elaborating on a one dimensional combined gradient-stochastic model and Tsallis q-statistics, as an illustrative example. The combined gradient-stochastic expression for the flow stress $\sigma$ reads

$$\sigma = \sigma^{ys} + h\varepsilon - \ell_p^2 \left( \partial^2 \varepsilon / \partial x^2 \right), \tag{5}$$

where the yield stress $\sigma^{ys}$ contains both an average and a fluctuating part given by $\sigma^{ys} = (1+\delta)\bar{\sigma}^{ys}$ – where $\bar{\sigma}^{ys}$ denotes mean value and $\delta$ follows a Weibull distribution fitted to experimental data. The rest of the quantities have their usual meanings; $\varepsilon$ is the strain; $h$ is a linear hardening modulus, and $\ell_p$ the deterministic internal length.

When this model is incorporated into a cellular automaton (CA) grid, it results to serrated stress-strain curves and power-law interpretations of the corresponding statistical events. Appropriate expressions for the stochastic component of the flow stress can be more fundamentally deduced by employing the formalism of random processes and stochastic differential equations [31,32]. In this connection, reference is made to a class of Tsallis q-distributions that are used in many non-equilibrium physics problems where the usual power-laws based on Boltzmann-Gibbs statistics fail to predict observed behavior. An expression used for interpreting deformation behavior of Mo micropillar compression is Tsallis q-exponential PDF of the form $P(s) = A[1 + (q-1)Bs]^{1/(1-q)}$: (A,B) are constants and the q-index is a measure of the system fractality, whereas $s$ denotes the burst size. A power-law relationship between the internal length $\ell_p$ and the entropic index q seems to hold, but this needs to be examined further [1].

Further elaboration along the above lines on combined gradient-stochastic models for the interpretation of size dependent serrated stress-strain graphs by employing Tsallis q-statistics and relate such findings with corresponding image observations on deformation patterns [1], is an open question that needs to be addressed in the future. In this connection, it is pointed out that the needed experimental information on appropriate PDF forms for the stochastic component of the flow stress can be deduced from multiple nanoindentation/NI tests and associated measurements of strain burst events. The PDF of the strain bursts would be related to a corresponding PDF for the flow stress on the assumption that a strain burst of a certain magnitude is the outcome of a number of material points yielding simultaneously. NI measurements at different locations and penetration depths can be conducted to deduce the statistical properties (mean, variance) for the local hardness which, in turn, one can extract information on the form of the stochastic component of the flow stress. From these multiple NI measurements one can extract direct information for both the deterministic ILs and the form of the stochastic contribution to the gradient dependent flow stress.

## 4.2 Chemomechanical Instabilities in LiB Anodes

In this subsection we provide elements of the ILG formulation that can be used to address chemomechanical instabilities in LiB anodes. In particular, we briefly present the basics of the stress-assisted diffusion and coupled ILG chemoelasticity theory that can be employed to consider local volume expansion in lithiated anodes and propagation of lithiation fronts.

*Size Dependent Stress-Assisted Diffusion:* The standard equations that are usually employed to model coupled elasto-diffusion processes are of the form

$$\sigma_{ij} = \lambda \varepsilon_{mm} \delta_{ij} + 2G\mu \varepsilon_{ij} - \alpha \rho \delta_{ij} \, , \ \mathbf{j} = -D\nabla \rho + M \rho \nabla \sigma_{ii} \qquad (6)$$

for the chemostress $\sigma_{ij}$ and the diffusive flux $\mathbf{j}$ where the coefficients $(\alpha, M)$ denote chemomechanical coupling constants and $D$ is the diffusivity. The fields $(\rho, \varepsilon_{ij})$ denote concentration of the diffusing chemical agent and mechanical strain, while $(\lambda, G)$ are the Lamé constants. Since these constitutive equations do not contain higher-order ILs, related chemomechanical size effects may not be captured.

Within our Laplacian-based ILG formulation, it turns out that the above constitutive equations are generalized by replacing $\sigma_{ij}$ with $\sigma_{ij} - \ell_\sigma^2 \nabla^2 \sigma_{ij}$; $\varepsilon_{ij}$ with $\varepsilon_{ij} - \ell_\varepsilon^2 \nabla^2 \varepsilon_{ij}$; and $\rho$ with $\rho - \ell_\rho^2 \nabla^2 \rho$, with $(\ell_\sigma / \ell_\varepsilon, \ell_\rho)$ denoting stress/strain and diffusional ILs. Under suitable assumptions, it is possible to uncouple the deformation and chemical fields by first computing a "ground" hydrostatic stress component $\sigma_h^0 = \sigma_{ii}^0$ from a conventional or gradient elasticity theory, and then derive the concentration $\rho$ from a stress-assisted diffusion equation of the form

$$\partial \rho / \partial t = \left( D + N\sigma_h^0 \right) \nabla^2 \left[ \rho - \ell_\rho^2 \nabla^2 \rho \right] - M \nabla \sigma_h^0 \cdot \nabla \left[ \rho - \ell_\rho^2 \nabla^2 \rho \right], \qquad (7)$$

where $N$ is a new phenomenological constant accounting for the effect of hydrostatic stress on diffusivity. This model with $\ell_\rho = 0$ has been used extensively to model hydrogen embrittlement and stress corrosion cracking in metals [37]. It can be adapted here, to consider chemomechanical damage and failure in LiB anodes.

*Size Dependent Lithiation Fronts:* To consider the propagation of lithiation fronts one may start with an expression for the free energy density $\psi$ of the form

$$\psi(\boldsymbol{\varepsilon}, \nabla e, \rho, \nabla \rho) = f(\rho) + \frac{1}{2} \kappa \nabla \rho \cdot \nabla \rho + \frac{1}{2} \boldsymbol{\varepsilon} \cdot \mathcal{C} \boldsymbol{\varepsilon} + \frac{1}{2} c \nabla e \cdot \nabla e, \qquad (8)$$

where $(\kappa, c)$ are respectively chemical and mechanical gradient coefficients, $\mathcal{C}$ denotes elasticity tensor, $\boldsymbol{\varepsilon}$ is the strain tensor and $e$ its hydrostatic part, while $\rho$ denotes concentration as before. In this case, both chemical ILs (through $\kappa$) and mechanical ILs (through $c$) enter into the formulation. Minimization of a corresponding energy functional yields field equations (and associated boundary conditions) for the local stress/strain and concentration of Li species, including the synergistic effect or interplay between higher-order mechanical and chemical ILs. The resulting coupled chemoelasticity governing equations for the stress and chemical species read

$$\boldsymbol{\sigma} = 2G\boldsymbol{\varepsilon} + \lambda(\mathrm{tr}\,\boldsymbol{\varepsilon})\mathbf{1} - \ell_x^2 \nabla^2 [2G\boldsymbol{\varepsilon} + \lambda(\mathrm{tr}\,\boldsymbol{\varepsilon})\mathbf{1}] - (2G + 3\lambda)M_o \rho \mathbf{1} \, , \qquad (9)$$

$$\mu = \mu^0 + RT\left[\ln\left(\frac{\rho}{1-\rho}\right) + \alpha(1-2\rho)\right] - \kappa\nabla^2\rho - \Omega_{Li}\sigma_h, \qquad (10)$$

where $(\lambda, G)$ are the Lamé constants, $R$ is the universal gas constant, $T$ is the absolute temperature, $\mu_0$ a reference value of the chemical potential, $\sigma_h = tr\boldsymbol{\sigma}/3$ the hydrostatic stress, and $\Omega_{Li} = 3M_0/c_{max}$ is the partial molar volume of the diffusing species.

### 4.3 Glioblastoma Instabilities in Brain

In this final third subsection we present some details on the GoG model, along with its mathematical similarities to the W-A model, and outline the potential new results to be expected from this comparison. Recent evidence in glioblastoma shows that one-size-fits-all vaso-modulatory interventions usually fail because control of glioma invasion characteristics, such as tumor front speed and infiltration width, vary widely and may require more personalized therapeutic interventions, in contrast to existing GoG models, which assume that all glioma cells have an identical GoG mechanism. In reality, each cell may have an idiosyncratic migration and proliferation regulation due to internal stress dependence and associated intrinsic heterogeneity. The relevant question is "how can we model and analyze the impact of such internal stress dependence and intrinsic heterogeneity of a tumor cell population, where migration and proliferation are regulated by the GoG mechanism".

This question can be addressed by incorporating internal stress effects in the D terms of the GoG model in analogy to the W-A model discussed earlier for structural defects. The GoG model as formulated by Hatzikirou and coworkers [17] reads

$$\partial\rho_m/\partial t = D_m\Delta\rho_m + E(\rho_m,\rho_i) \quad ; \quad \partial\rho_i/\partial t = D_i\Delta\rho_i - E(\rho_m,\rho_i) + g(\rho_i), \qquad (11)$$

where the $(\rho_m,\rho_i)$ denote respectively the motile and immotile glioma cell densities, with $(D_m, D_i)$ being the corresponding diffusion coefficients. The term $E(\rho_m,\rho_i)$ signifies the switching between the two different phenotypes. Finally, the function $g(\rho_i)$ denotes the cell proliferation of the immotile population. The phenomenological resemblance of the GoG model for motile-immotile cells to the W-A model for mobile-immobile dislocations is striking. The results obtained from the earlier study of the W-A model to consider heterogeneity, stochasticity and local stress dependence can be used to improve predictions of the GoG model. It is expected that these predictive results can enable to understand the impact of intratumoral heterogeneity in glioma progression: in particular, the persistence and size dependence of the Allee effect under different heterogeneity and internal stress distributions, as well as the role of the pertinent spatio-temporal instabilities on potential therapeutic failures.

## 5    ILG and Rheology: Newtonian and Complex Fluids

In this section we suggest possibilities for a gradient enhancement of constitutive equations used in fluid mechanics and rheology. In this connection, it is pointed out that following the author's work on gradient theory, a number of such generalizations have been proposed in these communities. For Newtonian fluids, such generalizations have been proposed by Silber and coworkers [38], as well as in more rigor and detail by Eliot and Gurtin [39]. For complex fluids, such generalizations can be found in the pioneering articles by Olmsted and coworkers (e.g. [40] and references quoted therein), as well as in the enlightening review by Cates and Fielding [41]; see also an earlier one by Dhont and Briels [42].

In the spirit of the ILG formulation such type generalizations can be readily deduced by replacing the local fields for the fluid density $\rho$, stretching tensor $\mathbf{D} = \frac{1}{2}[grad\,\mathbf{v} + (grad\,\mathbf{v})^T]$, and vorticity tensor $\mathbf{W} = \frac{1}{2}[grad\,\mathbf{v} - (grad\,\mathbf{v})^T]$ with their gradient-dependent counterparts $\rho - \ell_\rho^2 \nabla^2 \rho$, $\mathbf{D} - \ell_{\mathbf{D}}^2 \nabla^2 \mathbf{D}$, $\mathbf{W} - \ell_{\mathbf{W}}^2 \nabla^2 \mathbf{W}$.

Another possibility is to include the Laplacian of the viscoelastic stress $\mathbf{\Sigma}$ as proposed in the diffusive Johnson-Segalman (DJS) model employed to study shear banding flows of wormlike micelles or polymer solutions.

In such wormlike micellar systems, it is assumed [40] that the total stress $\mathbf{T}$ is separated into contributions from the Newtonian solvent and a viscoelastic stress $\mathbf{\Sigma}$ from the micelles, so that for creeping flow we have

$$\mathbf{T} = -p\mathbf{1} + 2\mu\mathbf{D} + \mathbf{\Sigma}; \; div\,\mathbf{T} = 0 \qquad (12)$$

with $p$ denoting the pressure, $\mu$ being the solvent's shear viscosity, and the second equation standing for quasi-static equilibrium. It is further assumed that the viscoelastic stress $\mathbf{\Sigma}$ obeys the following evolution equation

$$\overset{\circ}{\mathbf{\Sigma}} + \frac{1}{\tau}\mathbf{\Sigma} = \frac{2\mu^*}{\tau} + D\nabla^2\mathbf{\Sigma} , \qquad (13)$$

where $\tau$ denotes relaxation time, $\mu^*$ is the micelle polymer-like viscosity and $D$ is a diffusion-like coefficient. The corotational time derivative $\overset{\circ}{\mathbf{\Sigma}}$ may assumed to take various forms depending on the local micelle microstructural configuration. The above model and variants of it has been used extensively to model shear banding in complex fluids. The introduction of the Laplacian is needed to deal with ill- posedness in the negative slope regime of the shear stress – shear strain rate regime, i.e. the nonmonotonicity of the flow curve that also required the introduction of Laplacians in the author's gradint plasticity theory used to address shear banding in the deformation softening regime [3].

On returning to the topic of an appropriate generalization of the Navier-Stokes (N-S) equations for incompressible fluids, i.e. of the constitutive equation $\mathbf{T} = -p\mathbf{1} + 2\mu\mathbf{D}$, we can propose, in analogy to the gradient elasticity [7], the following gradient model

$$\mathbf{T} - \ell_{\mathbf{T}}^2 \nabla^2 \mathbf{T} = -p\mathbf{1} + 2\mu(\mathbf{D} - \ell_{\mathbf{D}}^2 \nabla^2 \mathbf{D}) , \qquad (14)$$

where $\ell_{\mathbf{T}}$ and $\ell_{\mathbf{D}}$ denote internal lengths associated with stress and strain rate inhomogeneities. On assuming that $\ell_{\mathbf{T}}$ can be neglected and introducing Eq. (14) in the equation of momentum balance $\rho\dot{\mathbf{v}} = div\,\mathbf{T}$ ($\rho$ is now the constant fluid density and $\dot{\mathbf{v}}$ its acceleration), we obtain the following gradient generalization of the N-S equations

$$\rho\dot{\mathbf{v}} = -\nabla p + \mu(\Delta\mathbf{v} - \ell_{\mathbf{D}}^2\Delta^2\mathbf{v}) , \qquad (15)$$

where $\Delta = \nabla^2$ and $\Delta^2 = \nabla^4$ denote the Laplacian and biharmonic or bi-Laplacian operators respectively. It is noted that Eq. (15) is identical to the equation used by Fried and Gurtin [39] to discuss plane Poiseuille liquid flow at small-length scales. A slightly generalized model was also used by the same authors to consider turbulence. The governing differential equations for this model read (in their notation)

$$\rho\dot{\mathbf{v}} = -\nabla p + \mu(1 - \alpha^2\Delta)\Delta\mathbf{v} + 2\rho\alpha^2 div\,\overset{\triangledown}{\mathbf{D}}, \qquad (16)$$

where the $\alpha$ parameter denotes a statistical correlation length and $\overset{\triangledown}{\mathbf{D}} = \overset{\bullet}{\mathbf{D}} + \mathbf{DW} - \mathbf{WD}$ denotes the usual corotational Jaumann rate.

Steady-state solutions of Eq. (16) with $\alpha = 0$, may be determined by employing the operator split method (or the use of Ru-Aifantis theorem [43]) utilized to eliminate singularities from dislocation lines and crack tips in the theory of gradient elasticity (see also [1]). This same procedure leads to the cancelation of singularities in typical fluid flow calculations involving immersed objects. It turns out, for example, that the resulting gradient Oseen tensor $\mathcal{O}_{ij}^G$, which generalizes its classical counterpart $\mathcal{O}_{ij}$

$$\mathcal{O}_{ij} = \frac{1}{8\pi\mu r}\left(\delta_{ij} + \frac{r_i r_j}{r^2}\right), \qquad (17)$$

where $r_i$ denotes the position vector and $r$ its magnitude, reads

$$\mathcal{O}_{ij}^G = \frac{1}{8\pi\mu r}\left\{\left[1 - 2e^{-r/\ell} - \frac{2\ell}{r}e^{-r/\ell} + \frac{2\ell^2}{r^2}(1 - e^{-r/\ell})\right]\delta_{ij} + \left[1 + 2e^{-r/\ell} + \frac{6\ell}{r}e^{-r/\ell} - \frac{6\ell^2}{r^2}(1 - e^{-r/\ell})\right]\frac{r_i r_j}{r^2}\right\}, \quad (18)$$

which resembles the exponential regularization of the Green's tensor in gradient elasticity and the resulting nonsingular gradient expressions for the stresses and strains in dislocation lines and crack tips. More details can be found in [44] where the authors seemed to be unaware of analogous developments in gradient elasticity.

## 6    ILG in Other Disciplines & Scales

In this section we summarize the applicability of the ILG framework to other disciplines and scales ranging from earth scales to quantum scales.

• *ILG in Geology:* Some initial work on introducing internal lengths and Laplacians of strain has been published by the author and coworkers to model shear banding and related instability phenomena in geomaterials including granular materials, soils, rocks and snow/ice (see, for example, [45-56]). Various types of gradient-dependent constitutive equations for such classes of geomaterials have also been proposed and elaborated upon in detail by many other authors. This was mainly due to the fact that the Laplacian was regularizing unstable behavior in the geomaterial's softening regime and allowed for the determination of shear band thickness and convergence of corresponding finite element calculations. The popularization of the approach in the geomechanics community is mainly due to the follow-up works by Vardoulakis and collaborators for soils, as well as de Borst and collaborators for concrete. These are too many to mention and can be found in the web.

In connection with the above, it is worth noting that the W-A model for dislocation patterning has recently been used by Ord and Hobbs [57] to interpret fracture patterns in frictional, cohesive, granular materials. Their article was one contribution of seventeen to a Theme Issue "Patterns in our planet: applications of multi-scale non-equilibrium thermodynamics to Earth-system science".

- *ILG in Electrodynamics:* The inclusion of higher-order gradients in deforming materials under the action of electromagnetic fields has also become very popular in recent years due to emerging applications and design of piezoelectric (induction of electricity due to applied pressure) and flexoelectric (induction of electricity due to strain gradients) components. The number of published articles is prohibitive to mention them here and we only refer to a couple of references by the author and coworkers [58-60], as well as the bibliography listed there for recent related literature on size effects.

In relation to the issue of eliminating singularities and introducing screening effects (e.g. Debye screening) in the electric and magnetic fields, the following gradient modification of Coulomb's law of electrostatics has been proposed (see, for example, [61] where a fractional generalization of Debye screening is also discussed)

$$\Delta \Phi(\mathbf{r}) - \frac{1}{r_D^2} \Phi(\mathbf{r}) = -\frac{1}{\varepsilon_0} \rho(\mathbf{r}), \qquad (19)$$

where $\Phi$ is the electrostatic potential [ $\mathbf{E}(\mathbf{r}) = -\nabla \Phi(\mathbf{r})$ ; $\mathbf{E}(\mathbf{r})$ is the electric field], $\rho(\mathbf{r})$ denotes now the charge density, $\varepsilon_0$ is the vacuum permittivity, $r_D$ is the Debye screening distance, and $\mathbf{r}$ denotes as usual the position vector. The classical Coulomb's potential for spherical symmetry at a point charge of strength $Q$ has the form $\Phi(\mathbf{r}) = Q / 4\pi\varepsilon_0 r$, while its Debye screened counterpart obtained from Eq. (19) (which is identical in form to the reduced Ru-Aifantis equation for gradient elasticity [43]) reads

$$\Phi(\mathbf{r}) = \frac{1}{4\pi\varepsilon_0} \frac{Q}{r} e^{-r/r_D}. \qquad (20)$$

In concluding this discussion on gradient electrodynamics, reference is made to an author's unpublished work where MacCullagh's 1850 proposal for an interesting formal analogy between elasticity and electromagnetism [62] is extended to include rotational gradients of the elastic aether. On assuming that the aether behaves as an elastic medium with its stress $\mathbf{T}$ depending linearly on rotations $\boldsymbol{\omega}$ (instead of strains), we have

$$\mathbf{T} = 2k\boldsymbol{\omega} \ , \ \ \boldsymbol{\omega} = \tfrac{1}{2}[\nabla \mathbf{u} - (\nabla \mathbf{u})^T], \ \ div\mathbf{T} = \rho \ddot{\mathbf{u}}, \qquad (21)$$

where $\mathbf{u}$ denotes displacement, $\rho$ density and $k$ an elastic constant. These lead to the equation $k\, curl\, curl\, \mathbf{u} + \rho \ddot{\mathbf{u}} = 0$ and by setting the terms $k\, curl\, \mathbf{u}$ and $\rho \dot{\mathbf{u}}$ to be proportional to the electric ($\mathbf{E}$) and magnetic ($\mathbf{B}$) fields respectively, we arrive at Maxwell's equations

$$\frac{\partial \mathbf{B}}{\partial t} + curl\, \mathbf{E} = 0 \ ; \ div\mathbf{B} = 0 \qquad \& \qquad \frac{\partial \mathbf{E}}{\partial t} - \frac{1}{\mu_0 \varepsilon_0} curl\, \mathbf{B} = 0 \ ; \ div\mathbf{E} = 0, \qquad (22)$$

where the identities $div\, curl\, \mathbf{u} = 0$ and $curl\,(\partial \mathbf{u} / \partial t) - \partial(curl\, \mathbf{u}) / \partial t = 0$, along with the following identification of the various coefficients ( $\beta = k\varepsilon_0$, $\mu_0 = \rho / \beta$ ; with $\beta$ being an arbitrary constant), were used. By adopting the above procedure, but replacing Eq. (21) for the aether's elastic stress with its gradient counterpart

$$\mathbf{T} = 2k(\boldsymbol{\omega} - \ell^2 \nabla^2 \boldsymbol{\omega}) \,, \tag{23}$$

we arrive at the following generalization of Maxwell's equations

$$\frac{\partial \mathbf{B}}{\partial t} + curl\left[(1 - \ell^2 \nabla^2)\mathbf{E}\right] = 0 \;\;;\;\; div\mathbf{B} = 0,$$
$$\frac{\partial \mathbf{E}}{\partial t} - \frac{1}{\mu_0 \varepsilon_0} curl\,\mathbf{B} = 0 \;\;\;\;\;\;;\;\;\;\; div\mathbf{E} = 0. \tag{24}$$

It is noted that for electrostatics under the assumption that the electric field $\mathbf{E}$ is proportional to a potential gradient $\nabla \Phi$, Podolsky's non-quantum electromagnetics[†] equation $\Delta\left[(1 - \ell^2 \Delta)\right]\Phi = 0$ is obtained.

- *ILG in Atomistics and Quantum Mechanics:* We conclude this section on applicability of the ILG framework to various disciplines and scales by focusing on two specific topics: A possible gradient generalization of the microscopic or molecular dynamics (MD) stress, and an analogous generalization of the quantum mechanical (QM) stress. In this connection, it is noted that the following expressions were proposed for these stresses:

$$\langle \boldsymbol{\sigma} \rangle = \frac{1}{V}\left[ \left\langle \frac{1}{2}\sum_i \mathbf{f}_{ij} \otimes (\mathbf{r}_i - \mathbf{r}_j) \right\rangle - \left\langle \sum_i m_i \mathbf{v}_i \otimes \mathbf{v}_i \right\rangle \right], \tag{25}$$

in [63], and

$$\sigma_{\alpha\beta} = -\frac{1}{V}\sum_i \left\langle \frac{p_{i\alpha} p_{i\beta}}{m_i} \right\rangle - \frac{1}{2V}\sum_{\substack{i,j \\ (j \neq i)}} \left\langle \frac{(\mathbf{r}_i - \mathbf{r}_j)_\alpha (\mathbf{r}_i - \mathbf{r}_j)_\beta}{|\mathbf{r}_i - \mathbf{r}_j|} U_{ij}^{'}\left(|\mathbf{r}_i - \mathbf{r}_j|\right) \right\rangle, \tag{26}$$

in [64], where the various symbols have their usual meaning. The striking formal similarity between these two expressions and their resemblance with the virial stress and other statistical stress measures is noted. However, the problem to connect such discrete "microscopic" stress measures with the continuum "macroscopic" measure of Cauchy stress in a "seamless" way is a challenging issue. A gradient generalization of the force fields $\mathbf{f}_{ij}$ in Eq. (25) and the interaction potential $U_{ij}$ in Eq. (26) may be appropriate which, among other things, could naturally introduce screening distances and eliminate associated singularities.

The effect of strain $\boldsymbol{\varepsilon}$ on the electronic structure has been described [65] through the equations

---

[†] Podolsky [Podolsky B. A generalized electrodynamics Part I—Non-quantum. *Phys. Rev.* (1942) **62**:68-71; Podolsky B., Schwed P. Review of a generalized electrodynamics. *Rev. Mod. Phys.* (1948) **20**:40-50] has derived a generalization of Maxwell's equations through a variational principle, leading into the appearance of $\nabla^2 \mathbf{B}$ in addition to $\nabla^2 \mathbf{E}$. This is also possible through the aforementioned analogy by replacing $\mathbf{u}$ with $\mathbf{u} - \ell^2 \nabla^2 \mathbf{u}$.

$$\left(E_c - \frac{\hbar^2}{2m^*}\nabla^2\right)\psi(\mathbf{r}) + a_c\, tr(\boldsymbol{\varepsilon})\psi(\mathbf{r}) = E\psi(\mathbf{r}) \quad \& \quad \boldsymbol{\varepsilon} = \boldsymbol{\varepsilon}^0; \; \boldsymbol{\sigma} = [\mathbf{C}]\boldsymbol{\varepsilon}\,; \;\; div\,\boldsymbol{\sigma} = \mathbf{0}, \qquad (27)$$

where $\psi(\mathbf{r})$ denotes the wavefunction, $\mathbf{C}$ is the Hookean elasticity matrix, $a_c$ the so-called deformation potential constant, and the rest of the symbols have their usual quantum mechanical meaning [65]. This is an uncoupled framework where strain can affect the electronic state but not vice-versa. A generalization to account for inverse effect on strain due to changes in the quantum field through the wavefunction $\psi(\mathbf{r})$, has already proposed as follows [66]:

$$\left(E_c - \frac{\hbar^2}{2m^*}\nabla^2\right)\psi(\mathbf{r}) + a_c\, tr(\boldsymbol{\varepsilon})\psi(\mathbf{r}) = E\psi(\mathbf{r}) \quad \& \quad \boldsymbol{\varepsilon} = \boldsymbol{\varepsilon}^0 - \frac{a_c}{3K}|\psi(\mathbf{r})|^2\,\mathbf{1}; \; \boldsymbol{\sigma} = [\mathbf{C}]\boldsymbol{\varepsilon}\,; \;\; div\,\boldsymbol{\sigma} = \mathbf{0}, \qquad (28)$$

where $K$ is the isotropic bulk elastic modulus. A possible gradient modification is then to replace $\boldsymbol{\varepsilon}$ with its gradient counterpart $\boldsymbol{\varepsilon} - \ell_\varepsilon^2\nabla^2\boldsymbol{\varepsilon}$, and this formal generalization may be of interest to further explore.

## 7    ILG Modification of Newton's Gradient Gravitation

In this section we venture a gradient generalization of Newton's Law which allows for the corresponding gravitational force to attain values larger than the electromagnetic force and even reach the levels of the nuclear and strong force which keeps matter together. The proposed modification is analogous to that earlier adopted by the author for gradient elasticity through the introduction of a Laplacian and a corresponding internal length[1†].

We begin with the following integral generalization of the gravitational force $\mathbf{f}$ in its component form ($f_i$):

$$f_i(\mathbf{r}) = \int G_{ij}(\mathbf{r} - \mathbf{r}')\,F_j(\mathbf{r}')\,d^3\mathbf{r}', \qquad (29)$$

where $G_{ij}(\mathbf{r} - \mathbf{r}')$ is a nonlocal interaction kernel and $F_j$ is the classical Newton's force. By assuming spherical symmetry/isotropy, Fourier transforming Eq. (29), Taylor series expanding up to the second order term, and inverting, we arrive at the following differential equation

$$\left(1 - \ell^2\nabla^2\right)\mathbf{f} = \mathbf{F} \quad ; \quad \ell^2\delta_{ij} = \frac{1}{2}\left|\frac{d^2\tilde{G}_{ij}(0)}{dk^2}\right|, \qquad (30)$$

where $k = |\mathbf{k}|$ denotes wave vector, $\tilde{G}_{ij}$ is the Fourier transform of $G_{ij}$ and $\ell$ is an internal length, with $\delta_{ij}$ appearing due to the assumed isotropy/spherical symmetry. In general, the sign in front of the Laplacian term of Eq. (30) may be positive or negative depending on the sign of $d^2\tilde{G}_{ij}(0)/dk^2$ of the second order term in the Taylor expansion. In other words, for $G_{ij} = \delta_{ij}G$ and

---

$\ell^2 = |l| = \left| d^2 \tilde{G}(k)/dk^2 \right|_{k=0}$, the term in the parenthesis of Eq. (30) becomes $\left(1 - \ell^2 \nabla^2\right)$ for $l > 0$ and $\left(1 - \text{sgn}[l]\ell^2 \nabla^2\right)$ for $l < 0$. Stability arguments may be employed to determine the sign of $l$ in a particular application.

Such a formal derivation can be also established by considering the two point masses $M_0$ and $M$ in the classical Newton's Law, as being distributed and bounded by spheres of finite radii. By considering, for example, the mass $M_0$ ($M_0 = \sum_i m_i$) being distributed within a sphere of radius $R_0$, summing up the interactions of each point mass $m_i$ (located at distance $r_i$ from the center of the sphere where $r_i = 0$) with the point mass $M$, and expanding in Taylor series the density $\rho(\mathbf{r}_i)$ around $\rho_0 = \rho(0)$ keeping terms up to the second order we obtain the following relationship

$$\mathbf{f} = \frac{GMV}{R^2}(\rho_0 + \ell^2 \nabla^2 \rho_0)\mathbf{e}_R \; ; \; \ell^2 = \frac{R_0^2}{10}, \tag{31}$$

where $\mathbf{e}_R = \mathbf{R}/R$ denotes the unit vector along the line connecting the center of $M_0$ with the point mass $M$. On setting $\int_V \rho_0 dV = M_0$, we then have $\mathbf{f} = (1 + \ell^2 \nabla^2)\mathbf{F}$ which by inversion leads to $(1 - \ell^2 \nabla^2)\mathbf{f} = \mathbf{F}$. This simplified rather intuitive calculation is similar to that earlier adopted by the author and coworkers (e.g. [10], [47]) to produce a corresponding gradient-dependent plastic strain.

On assuming a radial dependence of $\mathbf{f}$ and $\mathbf{F}$ [$\mathbf{f} = f(r)\mathbf{e}_r$ ; $\mathbf{F} = F\mathbf{e}_r, F = A/r^2$ with $A$ ($= GMM_0$) and G denoting now Newton's classical gravitational constant, where we have also adopted the notation $\mathbf{e}_R \equiv \mathbf{e}_r$], we can readily solve the scalar counterpart of $(1 - \ell^2 \nabla^2)\mathbf{f} = \mathbf{F}$, i.e.

$$f - \ell^2 \left(\frac{\partial^2 f}{\partial r^2} + \frac{2}{r}\frac{\partial f}{\partial r} - \frac{2f}{r^2}\right) = \frac{A}{r^2}, \tag{32}$$

by also requiring that $\mathbf{F} \to 0$ as $r \to \infty$. The result is

$$f = \frac{A}{r^2}[1 + Be^{-r/\ell}(1 + \frac{r}{\ell})], \tag{33}$$

where $B$ is a new parameter to evaluate in connection with experiments. It is noted that the above expression of Eq. (33) reduces to Newton's classical force $F_N = A/r^2$ as $r \to \infty$ and to the expression $F_{SF} = AB/r^2$ as $r \to 0$. By adjusting the value of the new parameter $B$ ($B \gg 1$) we can attain values of the nuclear and strong force.

The internal length parameter can be identified with the de Broglie relativistic length, the Compton length, the Planck length or the Schwarzschild distance, according to the configuration at hand, i.e.

- De Broglie: $\ell = \hbar / \gamma m_0 c$ …………..$6.309 \times 10^{-16}\,m$ ,
- Compton: $\ell = \hbar / m_p c$ …………....$2.10 \times 10^{-16}\,m$ ,
- Planck: $\ell = \sqrt{\hbar G / c^3}$ ……………..$1.616 \times 10^{-35}\,m$ ,
- Schwarzschild: $\ell = 2Gm_{BH}/c^2 \ldots 10^{10} - 10^{13}\,m$ ,

where $\hbar$ denotes the Planck constant, $c$ is the speed of light; and $(m_0, m_p, m_{BH})$ denote rest masses for neutrino, proton, and black hole, respectively; whereas $G$ in the above denotes the classical Newton's gravitational constant (not to be confused with the same symbol earlier used for the shear modulus), and $\gamma$ is the Lorentz factor ($\gamma = 1/\sqrt{1-(v/c)^2}$; with $v$ denoting particle speed), not to be confused with a similar symbol used in earlier sections for the strain.

On adopting the Vayenas and coworkers Rotating Neutrino model (RNM) for the nucleus [67,68] we now utilize the above expression for the gravitational force given by Eq. (33), in conjunction with the centrifugal force $F_C = \gamma m_0 c^2 / r$, where $r$ denotes the radius of the nucleus modeled by the three rotating neutrinos whose total relativistic mass is $m_N = 3\gamma m_0$. An estimate of $\gamma$ can be obtained by equating the proton energy $m_p c^2$ with the relativistic neutrino mass. This gives the value of $\gamma = m_p / m_N$ which, according to experimental measurements for $m_p$ and $m_0$ turns out to be equal to $7.818 \times 10^9$.

Having such a value of $\gamma$ available, we can make effective use of the aforementioned equality between gravitational and centrifugal forces in Vayenas' RNM to deduce the relationship

$$F_R = \frac{A}{\sqrt{3}r^2}[1 + Be^{-\sqrt{3}r/\ell}(1 + \frac{\sqrt{3}r}{\ell})] = F_C = \frac{\gamma m_0 c^2}{r}, \qquad (34)$$

where the factor $\sqrt{3}$ rises by considering the resultant gravitational force $F_R = \sqrt{3}f$ due to the interaction of the 3 symmetrically placed (at angles $120°$) rotated neutrinos. One possibility for the constant $A$ being set it equal to $A = Gm_0^2\gamma^2$, to account for relativistic effects during the interaction of each pair of neutrinos in the assumed RNM configuration. The above relationship (for $\ell$ identified with de Broglie's relativistic length $\ell = \hbar / \gamma m_0 c = 6.31 \times 10^{-16} m$) gives the following value for the coefficient $B$

$$B = \frac{\sqrt{3}e^{\sqrt{3}}\hbar c}{(1+\sqrt{3})Gm_0^2\gamma^2} = 5.47 \times 10^{39}, \qquad (35)$$

and a corresponding value of $F_R$

$$F_R = 7.92 \times 10^4 \, \text{N}, \qquad (36)$$

i.e. the value of the strong force obtained for the RNM configuration [67,68] by using an entirely different approach. In that approach Eq. (34) with $B = 0$ was used with $A = Gm_p^2\gamma^6$ giving a value for $\gamma = 3^{1/12}m_{pl}^{1/3}m_0^{-1/3} = 7.167 \times 10^9$, where $m_{pl}$ is the Planck mass ($m_{pl} = \sqrt{\hbar c/G}$), and the value of $m_0$ was taken as $m_0 = 0.0436 \, eV/c^2$. And since $\gamma = m_p/3m_0$, this gives $m_p = 9.38 \times 10^8 \, eV/c^2$, i.e. the same value as the one used in the previous paragraph by properly adjusting the parameters $(A, B)$, as well as by identifying the internal length parameter $\ell$ with de Broglie relativistic length. Other choices of $(A, B, \ell)$ are possible not only for the RNM configuration at hand, but also other more complex geometric models for elementary particles represented by several neutrinos where a potential is convenient to use.

## 8    *Gradient Interatomic Potentials*

Motivated by the above extension of Newton's gravitational potential, we consider in this section a similar gradient generalization of London's quantum mechanical potential. Based on exact quantum mechanical calculations London [69,70] has arrived at the following forms of the interatomic force $F\left(=-dw\,/\,dr\right)$ and interatomic potential $w\left(=w(r)\right)$

$$F=-\frac{dw}{dr}\;;\;\; w=w(r)=\begin{cases}-\dfrac{3\alpha_0^2 hv}{4(4\pi\varepsilon_0)^2}\dfrac{1}{r^6}=-\dfrac{C}{r^6};\;\; r\geq\sigma,\\[4mm] \infty;\;\; r<\sigma,\end{cases}\tag{37}$$

where $C=3\alpha_0^2 hv\,/\,4(4\pi\varepsilon_0)^2$, $\alpha_0$ is the atomic polarizability and $\varepsilon_0$ the vacuum dielectric permittivity. The quantities $\left(h,v\right)$ denote respectively the Planck constant ($h=2\pi\hbar$) and the electron orbital frequency. It is noted that the above form provides an explicit expression for the attractive interaction until a critical distance $\sigma$ below which the model breaks down as the interaction becomes repulsive going to infinity as $r\to 0$. To describe quantitatively "repulsive" interactions for distances $r<\sigma$, Lennard-Jones [71] suggested the following modification of London's potential

$$w_{L-J}(r)=-\frac{A}{r^6}+\frac{B}{r^{12}},\tag{38}$$

where $A$ and $B$ are determined by fitting them to obtain through atomistic simulations the measured experimental values of macroscopic properties. Other type of interaction potentials can be found in [72].

The gradient modification of London's quantum mechanical potential, denoted as $w_L^G$, is obtained in terms of its classical counterpart $w$ through the inhomogeneous Helmholtz equation

$$\left(1-\ell^2\nabla^2\right)w_L^G=w,\;\; w=w(r)=-\frac{C}{r^6}\;.\tag{39}$$

The solution of Eq. (39) for ($w_L^G\to 0, r\to\infty$) is given by the expression

$$w_L^G(r)=A\ell\frac{e^{-r/\ell}}{r}+\frac{C}{48\ell^6}\left\{\frac{4\ell^4}{r^4}+\frac{2\ell^2}{r^2}+\frac{\ell}{r}\left[e^{r/\ell}\,\text{Ei}(-\frac{r}{\ell})-e^{-r/\ell}\,\text{Ei}(\frac{r}{\ell})\right]\right\},\tag{40}$$

where $A$ is a new integration constant, $\ell$ is an internal length parameter, and $\text{Ei}$ denotes the exponential integral $\text{Ei}(x)=-\int_{-x}^{\infty}\frac{e^{-t}}{t}dt$. Near the origin ($r\to 0$), it turns out that $w_L^G(r)\to\frac{C}{12\ell^2 r^4}$, while at large distances ($r\to\infty$) it approaches the classical London's potential, i.e. $w_L^G(r)\to-\frac{C}{r^6}$ for $r>>\ell$.

As an example application of the newly derived gradient potential, we consider the case of Argon (Ar). It has been shown that the Lennard-Jones potential is able to describe accurately the simulated liquid argon properties in agreement with the experiment. Numerical/experimental values can be utilized by the data provided in [73] (see also Table 6.1 of [72]). Among these data, of particular interest are the minimum of the potential function, designated as $\varepsilon$ (in units of Joules or eV), as well as its location $r_m$ (in $\overset{\circ}{A}$). Their estimated values are $\varepsilon = 1.95 \times 10^{-21} J$ and $r_m = 0.37 nm$ respectively. The Lennard-Jones potential can be uniquely determined from these parameters. For this purpose, Eq. (38) is written in the form $w_{L-J}(r) = \varepsilon((r_m/r)^{12} - 2(r_m/r)^6)$, where it is evident that the minimum occurs at $r_m$ with $w_{L-J}(r_m) = -\varepsilon$ and $dw_{L-J}(r_m)/dr = 0$. This point determines the transition from "attractive" to "repulsive" branch for distances $r < r_m$. Additionally, the Lennard-Jones potential is zero at $r = \sigma = 2^{-1/6} r_m = 0.89 r_m = 0.32 nm$. The parameters $(\varepsilon, r_m)$ are related with the $(A, B)$ of Eq. (38) through the relationship $A = \varepsilon r_m^{12} = 4\varepsilon\sigma^{12}$, $B = 2\varepsilon r_m^6 = 4\varepsilon\sigma^6$. The fitted London's constant is $C = 50 \times 10^{-79} J \cdot m^6$, which was determined such as the classical London's potential passes through the experimental potential minimum exactly at $r_m$.

In order to demonstrate the ability of the gradient modification of London's potential to recover the behavior of the Lennard-Jones potential for the Ar-Ar interaction case, we perform adjustment of the gradient parameters $(A, \ell, C)$, such as the position of the potential minimum occurs at $r_m$, i.e. $w_L^G(r_m) = -\varepsilon$, and the corresponding potential curves are as close as possible by minimizing their mean square error. The obtained parameter values are $A = 1.392 \times 10^{-17} J$, $\ell = 0.57 \overset{\circ}{A}$, and $C = 90 \times 10^{-79} J \cdot m^6$. The estimated values for the internal length are consistent with the atomistic simulations. As shown in Fig. 1a, the gradient modification of London's potential fits nicely the Lennard-Jones, while both empirical curves have their minima intersect at distance $r_m$. It is noted that the gradient potential has the same asymptotic $O(r^{-6})$ distances $r > r_m$, in agreement with both Lennard-Jones and London's potential. Finally, as expected, the gradient modified London's potential becomes "repulsive" for $r < r_m$, where the change of slope occurs, in contrast to London's original $1/r^6$ monotonic potential.

Another indicative example of the applicability of the newly derived gradient potential is the Stillinger-Weber potential, which is broadly used to model the interatomic interactions of materials with diamond structures such as crystalline semiconductors (Si, Ge). The analytical expression of the two-body Stillinger-Weber potential reads [74]

$$w_{S-W}(r) = \begin{cases} \varepsilon A \left[ B \dfrac{\sigma^4}{r^4} - 1 \right] \exp\left( \dfrac{1}{r/\sigma - a} \right); & r < a\sigma, \\ 0; & r \geq a\sigma. \end{cases} \tag{41}$$

The suggested fitted values for the Stillinger-Weber potential when applied to Si semiconductor read $A = 7.04955627$, $B = 0.602224558$, $\varepsilon = 50 kcal \times mol^{-1} = 3.4723 \times 10^{-19} J$, $\sigma = 2.0951 \overset{\circ}{A}$, and $a = 1.8$ [19]. The Stillinger-Weber potential has a cutoff at distance $r = a\sigma$, confining the interatomic interaction within that range, while for short distances it has a "repulsive" branch with asymptotic behavior $w_{S-W}(r) \to \varepsilon e^{-1/a} A B \sigma^4 / r^4$.

The estimated values for the gradient potential are $A = 5.702 \times 10^{-17} J$, $\ell = 0.974 \overset{\circ}{A}$, and $C = 1.333 \times 10^{-78} J \cdot m^6$ respectively. They are adjusted such as the fitted minimum of $w_L^G$ coincides with the corresponding one of the Stillinger-Weber potential, which satisfies $w_{S-W}(r_m) = -\varepsilon$ at $r_m = 1.118\sigma = 2.34 \overset{o}{A}$. In Figure 1b, it is demonstrated that the parameters $(A, \ell, C)$ can be further adjusted in order to describe accurately the behavior of $w_{S-W}$ for small distances. This is possible due to the fact that $w_L^G(r) \rightarrow \dfrac{C}{12\ell^2 r^4}$ as $r \rightarrow 0$, which is in agreement with the asymptotic behavior of $w_{S-W}$ at the origin.

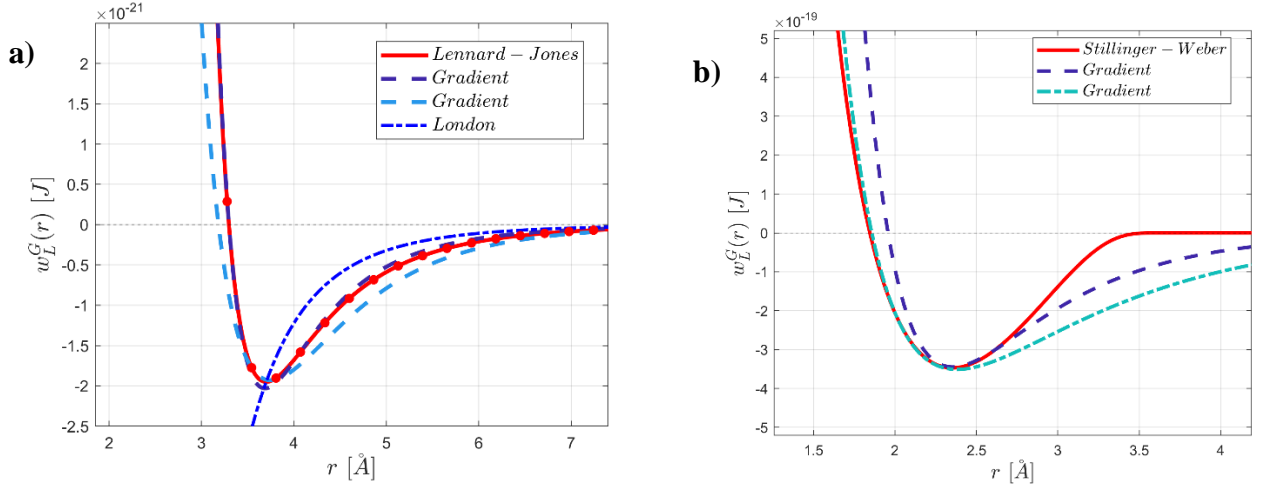

**Figure 1**: Quantitative plots of the gradient London potential fitting to **a)** Lennard-Jones (Ar-Ar) and **b)** Stillinger-Weber (Si) potential respectively.

Another possible generalization of the gradient approach for constructing new interatomic potentials is through the introduction of an additional biharmonic or bi-Laplacian term in London's potential. Motivated by the 4$^{th}$ order GradEla extension used earlier by Aifantis and co-workers [75-77] we can further generalize Eq. (39) to read

$$\left(1 - \tilde{\ell}_1^2 \nabla^2 + \tilde{\ell}_2^4 \nabla^4\right) w_L^G = w_L(r), \qquad (42)$$

where $\left(\tilde{\ell}_1, \tilde{\ell}_2\right)$ now denote two internal lengths. In passing, it is noted that such a fourth-order equation for the elastic strain, derived within a second strain gradient elasticity/GradEla theory [75-79], leads to the elimination of singularities of the dislocation density tensor, which remains singular in first strain GradEla [77] .

Equation (42) can be factored as the product of two Helmholtz operators as $\left(1 - \ell_1^2 \nabla^2\right)\left(1 - \ell_2^2 \nabla^2\right) w_L^G = w_L(r)$, where the internal lengths $\left(\ell_1, \ell_2\right)$ are given by the expression $\ell_{1,2}^2 = \tilde{\ell}_1^2 (1 \pm \sqrt{1 - 4\dfrac{\tilde{\ell}_2^4}{\tilde{\ell}_1^4}})$. The solution of Eq. (42) is obtained by applying the operator split approach of Ru-Aifantis theorem, arriving at the equation $\left(1 - \ell_2^2 \nabla^2\right) w_L^G = w_L^{G,1}\left(r; \ell_1\right)$, where $w_L^{G,1}\left(r; \ell_1\right)$ is the gradient London's potential of Eq. (40) with internal length $\ell_1$. It is given by the expression

$$w_L^G(r) = \frac{B\,\ell_2^3}{\ell_2^2 - \ell_1^2}\frac{e^{-r/\ell_2}}{r} + \frac{A\,\ell_1^3}{\ell_1^2 - \ell_2^2}\frac{e^{-r/\ell_1}}{r} - \frac{C}{48\ell_1^3\ell_2^3}\left\{ \frac{\ell_2^3}{(\ell_1^2 - \ell_2^2)r}\left[ e^{r/\ell_1}\,\mathrm{Ei}(-\frac{r}{\ell_1}) - e^{-r/\ell_1}\,\mathrm{Ei}(\frac{r}{\ell_1}) \right] \right.$$

$$\left. + \frac{2\ell_1^3\ell_2}{(\ell_1^2 - \ell_2^2)r^2} + \frac{2\ell_2^3\ell_2}{(\ell_2^2 - \ell_1^2)r^2} + \frac{\ell_1^3}{(\ell_2^2 - \ell_1^2)r}\left[ e^{r/\ell_2}\,\mathrm{Ei}(-\frac{r}{\ell_2}) - e^{-r/\ell_2}\,\mathrm{Ei}(\frac{r}{\ell_2}) \right] \right\}. \quad (43)$$

where $B$ is a new integration constant and $(\ell_1, \ell_2)$ have been defined above, while $C$ is the London's constant and $A$ the integration constant of Eq. (40). It is noted that the first two exponential terms in Eq. (43), which are related to the homogeneous part of their corresponding Helmholtz equations, are formally similar with the expressions derived earlier for nuclear potentials in quantum electrodynamics based on an extension of Yukawa-type interactions. A quantitative elaboration for specific material types will be a subject of a future study[†].

## 9 Fractional Considerations

In this final section an extension of the ILG framework to incorporate fractional derivatives is presented. A fractional generalization of GradEla is established by replacing the standard (integer) Laplacian $\Delta \equiv \nabla^2$ with a fractional one of the Riesz form $(-\Delta)^{\alpha/2}$ (or the Caputo form $^C\Delta_w^\alpha$) in the constitutive expression $\sigma_{ij} = \lambda\varepsilon_{kk}\delta_{ij} + 2G\varepsilon_{ij} - \ell^2\nabla^2[\lambda\varepsilon_{kk}\delta_{ij} + 2G\varepsilon_{ij}]$. An example of such a fractional generalization reads [78]

$$\sigma_{ij} = (\lambda\varepsilon_{kk}\delta_{ij} + 2G\varepsilon_{ij}) + \ell^\alpha (-\Delta)^{\alpha/2}[\lambda\varepsilon_{kk}\delta_{ij} + 2G\varepsilon_{ij}], \quad (44)$$

where $(-\Delta)^{\alpha/2}$ is the fractional generalization of the Laplacian in the Riesz form, defined in terms of the Fourier transform $\mathcal{F}$ by

$$\left((-\Delta)^{\alpha/2}\varepsilon_{ij}\right)(\mathbf{r}) = \mathcal{F}^{-1}\left( |\mathbf{k}|^\alpha\,\tilde{\varepsilon}_{ij}(\mathbf{k}) \right)(\mathbf{r}), \quad (45)$$

where $\mathbf{k}$ denotes the wave vector, and $\tilde{\varepsilon}_{ij}(\mathbf{k}) = \mathcal{F}\left(\varepsilon_{ij}(\mathbf{r})\right)(\mathbf{k})$ is the Fourier transform of $\varepsilon_{ij}(\mathbf{r})$. On introducing the fractional GradEla constitutive relation given by Eq. (44) into the equilibrium relation $div\boldsymbol{\sigma} = 0$ we obtain

$$[1 + \ell^\alpha (-\Delta)^{\alpha/2}][\lambda\nabla tr\,\boldsymbol{\varepsilon} + 2G\,div\,\boldsymbol{\varepsilon}] = 0, \quad (46)$$

where the notation $\ell^\alpha$ is adopted for the corresponding fractional internal length. Noting the fact that the operators $\nabla$ and $(-\Delta)^{\alpha/2}$ commute and that the second bracket in Eq. (46) is also zero by replacing $\boldsymbol{\varepsilon}$ with $\boldsymbol{\varepsilon}_0$, where $\boldsymbol{\varepsilon}_0$ denotes the solution of the corresponding equation for classical elasticity, (i.e. $\lambda \nabla tr \, \boldsymbol{\varepsilon}_0 + 2G \, div \, \boldsymbol{\varepsilon}_0 = 0$ ), we can easily deduce that the solution of Eq. (46) satisfies the reduced fractional partial differential equation

$$[1 + \ell^\alpha (-\Delta)^{\alpha/2}] \boldsymbol{\varepsilon} = \boldsymbol{\varepsilon}_0 \,, \tag{47}$$

which for the case α = 2 reduces to the inhomogeneous Helmholtz equation. It is thus critical to derive fundamental solutions for Eq. (47); i.e. for the equation

$$[1 + \ell^\alpha (-\Delta)^{\alpha/2}] G_\alpha(\mathbf{r}) = \delta(\mathbf{r}) \,, \tag{48}$$

where $G_\alpha(\mathbf{r})$ denotes the fundamental solution, $\delta(\mathbf{r})$ denotes the delta function and $\mathbf{r}$ is the radial coordinate in a 3D space. To compute the fundamental solution of Eq. (48) with the natural boundary condition $G_\alpha(\mathbf{r}) \to 0$ as $\mathbf{r} \to \infty$, we employ the method of Fourier transforms. Using the properties of the Fourier transform of the Riesz fractional Laplacian as defined by Eq. (45), along with the well-known transform of the delta function $\mathcal{F}\{\delta(\mathbf{r})\}(\mathbf{k}) = 1$, we obtain the algebraic equation $\left(1 + \ell^\alpha |\mathbf{k}|^\alpha\right) G_\alpha(\mathbf{k}) = 1$, which gives for the fundamental solution $G_\alpha(\mathbf{k}) = \left(1 + \ell^\alpha |\mathbf{k}|^\alpha\right)^{-1}$. Consequently, the fundamental solution of Eq. (48) in the physical space is obtained through inversion as

$$G_\alpha(\mathbf{r}) = \frac{1}{(2\pi)^3} \int \frac{1}{1 + \ell^\alpha |\mathbf{k}|^\alpha} e^{i\mathbf{k}\cdot\mathbf{r}} d^3\mathbf{k} \,. \tag{49}$$

---

[†] For completeness, however, we may refer to the paper by Reid [Reid R.V. Local phenomenological nucleon–nucleon potentials, *Annals of Physics* (1968) **50**:411-448], where the following expression, among others, is proposed $V_C = h\left[ e^{-x/3} - 13.8\,e^{-3x} + 138\,e^{-6x} \right]/x$, where $h = 10.463\ MeV$ and $x = \mu r, \mu = mc/\hbar \approx 0.7\ \mathrm{fm}^{-1}$.

The inversion of Eq. (49) is performed through application of the convolution property of the Mellin transform, along with a corresponding Mellin-Barnes integral representation, which yields the following corresponding Fox-H function expression [78$^c$]

$$G_\alpha(\mathbf{r}) = \frac{1}{2\alpha \, \pi^{3/2} \ell^{-1} |\mathbf{r}|^2} H_{1,3}^{2,1} \left[ \frac{|\mathbf{r}|}{2\ell} ; \begin{array}{cc} (1 - \frac{1}{\alpha}, \frac{1}{\alpha}) \\ (1 - \frac{1}{\alpha}, \frac{1}{\alpha}) & (1, \frac{1}{2}) \quad (\frac{1}{2}, \frac{1}{2}) \end{array} \right]. \tag{50}$$

For more details concerning the definition, properties and applications of the Fox-H function in fractional analysis, the reader can consult [79]. A corresponding series expansion of Eq. (50) is also provided in [78$^c$]. It is noted that as $\alpha \to 2$, we obtain the Green's function of the integer order Helmholtz equation, i.e. $G_\alpha(\mathbf{r})|_{\alpha \to 2} = (1/4\pi) e^{-r/\ell}$.

Motivated by the above analysis of the fractional Helmholtz equation, as well as by noticing that Eq. (49) reduces to a Yukawa-like potential in the classical limit $\alpha \to 2$, a fractional treatment of Eq. (29), listed below again for convenience,

$$f_i(\mathbf{r}) = \int K_{ij}(\mathbf{r} - \mathbf{r}') F_j(\mathbf{r}') d^3\mathbf{r}' \,, \tag{51}$$

is undertaken, where now $K_{ij}$ denotes a fractional interaction kernel, and the forces $f_i$, $F_j$ have been defined in Section 7. Such type of integral expressions have been previously introduced to model nonlocal constitutive relations in electrodynamics leading to fractional Debye screening effects [61].

Similar arguments have been recently applied to model fractional nonlocality in GradEla [78[d]]. Through a Taylor series expansion (up to second order) of the fractional kernel $K_{ij}$ in Fourier space involving non-integer powers of the wave vector $|\mathbf{k}|^{\alpha}$, and subsequent inversion through Eq. (45), the corresponding fractional counterpart of Eq. (30) is obtained as

$$\left(1+\ell^{\alpha}\left(-\Delta\right)^{\alpha/2}\right)\mathbf{f}=\mathbf{F} \quad ; \quad \ell^{\alpha}\delta_{ij}=\frac{1}{\Gamma\left(\alpha+1\right)}\left|\left(_{0}^{C}D_{k}^{\alpha}\tilde{K}_{ij}\right)(0)\right| ,$$ (52)

where $_{0}^{C}D_{k}^{\alpha}\tilde{K}_{ij}$ is the Caputo fractional derivative of order $\alpha$ with respect to $k$ [79]. In the limit $\alpha \to 2$ the solution Eq. (52) coincides with the one obtained from Eq. (30), since the fractional Laplacian and corresponding derivatives reduce to their classical counterparts. The solution of Eq. (52) can be obtained through convolution of the corresponding Green's function of Eq.(49) with the classical field $\mathbf{F}$, i.e. $f_i = G_{\alpha}*F_i$, resulting to the expression

$$f = \frac{A}{r^2}[1+BK_{\alpha}(r/\ell)],$$ (53)

where $A, B$ are the constants, defined in Section 7, and $K_{\alpha}$ is the fractional generalization of the modified (fractional-like) Bessel function $K_{\alpha}\left(r\right) \equiv \frac{1}{2\pi^2 r}\int_{0}^{\infty}\frac{k^2\cos(k\,r)}{k^{2-\alpha}\left(1+\ell^{\alpha}k^{\alpha}\right)}\,dk$. An analogous result can be obtained by further generalizing Eqs. (30) and (52) to include bi-Laplacian terms of the type appearing in Eq. (42) for both the integer and non-integer case. This will be a subject of future publication. However, for the completeness of this review, we list the corresponding formulas below

$$f = \frac{A}{r^2}[1+B_1 e^{-r/\ell_1}(1+\frac{r}{\ell_1})+B_2 e^{-r/\ell_2}(1+\frac{r}{\ell_2})] \quad \text{integer case,}$$ (54)

$$f = \frac{A}{r^2}[1+B_1 K_{\alpha}(r/\ell_1)+B_2 K_{\alpha}(r/\ell_2)] \quad \text{non-integer case,}$$ (55)

where $B_1, B_2$ are dimensionless constants, $(\ell_1, \ell_2)$ are two internal lengths, while the function $K_{\alpha}\left(r\right)$ has been defined above.

## Conclusions

A concise review of gradient models (across scales, materials, and processes) was provided based on the author's ILG approach. As a result, earlier references on generalized continuum mechanics and recent contributions on gradient and non-local theories were not discussed due to space limitation. For solids, one should single out the contributions of Eringen [80], Fleck and Hutchinson [81] Gurtin and Anand [82], Gao et al [83[a]], Nix and Gao [83[b]], de Borst et al. [84], Geers et al [85[a]], Peerlings et al [85[b]], Willis et al [86[a]], Aifantis and Willis [86[b]], and Polizzotto [87]. Many more are included in a most recent and detailed article by Voyiadjis and Song [88] focusing on gradient plasticity. Gradients in fluid and granular flows were considered most recently by Goddard [89]. For additional recent developments on granular flow, one may also consult references [90–93], while for internal length interpretations based on kinetic theory, one may consult [94]. However, the intention of the article was not to elaborate on and review the various important classical-like gradient models for solids and fluids, as well as for rheology and electrodynamics. Its main purpose was to explore the applicability of gradient theory for scales and processes not considered before, and point out its potential usefulness for atomistic simulations and elementary particles, as well as for earth and planetary processes. In this connection, it is noted that while completing this article, it came to the attention of the author that an expression similar to that derived herein and given by Eq. (33) was also proposed on rather intuitive grounds by Fischbach et al [95$^a$] in an effort to re-interpret existing measurements on earth's gravity (see also [95$^b$]). The values of their constants were entirely different than ours, as they used it for a reanalysis of the Eötvös experiment on Earth's gravitational field. There has been a vast literature on this expression, subsequently referred to as the "fifth force," which we will discuss in a forthcoming publication, as this is beyond the scope of the present review.

## Acknowledgements

The work was greatly benefited from the RISE/FRAMED project no. 734485 (https://cordis.europa.eu/project/rcn/207050/factsheet/en) for which Aristotle University of Thessaloniki (AUTh) acts as coordinator. In this connection, thanks are extended to all beneficiary nodes and international partners of FRAMED. The gradient fluids section is a topic of the RISE/ATM2BT project no. 824022 (https://cordis.europa.eu/project/rcn/219192/factsheet/en) for which EU support is also acknowledged. This section was included in anticipation of follow-up joint work between AUTh, Akita Univeristy/Japan and Aston University/UK (which acts as project's ATM2BT coordinator). The remaining of the sections were benefited from discussions with C. Vayenas, K.E. Aifantis and K. Parisis.

## References

1. Aifantis, E.C. Internal length gradient (ILG) material mechanics across scales and disciplines. *Adv. Appl. Mech.* (2016) **49**:1–110.

2. Aifantis, E.C. On the Microstructural Origin of Certain Inelastic Models. *J. Eng. Mater. Technol.* (1984) **106**:326–330.

3. Aifantis, E.C. The physics of plastic deformation. *Int. J. Plast.* (1987) **3**:211–247.

4. Aifantis, E.C. On the role of gradients in the localization of deformation and fracture. *Int. J. Eng. Sci.* (1992) **30**:1279–1299.

5. Aifantis, E.C. Pattern formation in plasticity. *Int. J. Eng. Sci.* (1995) **33**:2161–2178.

6. Aifantis, E.C. On scale invariance in anisotropic plasticity, gradient plasticity and gradient elasticity. *Int. J. Eng. Sci.* (2009) **47**:1089-1099.

7. Aifantis, E.C. On the gradient approach – Relation to Eringen's nonlocal theory. *Int. J. Eng. Sci.* (2011) **49**:1367–1377.

8. Aifantis, E.C. Gradient Nanomechanics: Applications to Deformation, Fracture, and Diffusion in Nanopolycrystals. *Metall. Mater. Trans. A.* (2011) **42**:2985–2998.

9. Askes, H., Aifantis, E.C. Gradient elasticity in statics and dynamics: An overview of formulations, length scale identification procedures, finite element implementations and new results. *Int. J. Solids Struct.* (2011) **48**:1962–1990.

10. Aifantis, E.C. Gradient material mechanics: Perspectives and Prospects. *Acta Mech.* (2014) **225**: 999–1012.

11. [a]Aifantis E.C., Serrin J. B. The mechanical theory of fluid interfaces and Maxwell's rule, *J. Coll. Inter. Sci.* (1983) **96**:517-529; [b]Aifantis E.C., Serrin J.B. Equilibrium solutions in the mechanical theory of fluid microstructures, *J. Coll. Inter. Sci.* (1983) **96**:530-547; [c]Van der Waals J.D. Théorie thermodynamique de la capillarité, dans l'hypothèse d'une variation continue de densité, *Arch. Neerl. Sci. Exactes Nat.* (1895) **28**:121-209; [d]D. ter Haar (Editor), *Collected papers of L.D. Landau*, Pergamon, London, 1965; [e]Cahn J.W., Hilliard J.E., Free energy of a nonuniform system. I. Interfacial free energy, *J. Chem. Phys.* (1958) **28**:258-267; [f]Cahn J.W. Free energy of a nonuniform system. II. Thermodynamic basis, *J. Chem. Phys.* (1959) **30**:1121-1124.

12. [a]Kevrekidis I.G., Gear C.W., Hyman J.M., Kevrekidis P.J., Runborg O., Theodoropoulos C. Equation-free, coarse-grained multiscale computation: Enabling macroscopic simulators to perform system-level analysis, *Comm. Math. Sci.* (2003 ) **1** :715-762; [b]Kevrekidis I.G., Samaey G. Equation-Free Multiscale Computation: Algorithms and Applications, *Annu. Rev. Phys. Chem.* (2009) **60**:321-344.

13. [a]Tsallis C. Possible generalization of Boltzmann-Gibbs statistics, *J. Stat. Phys.* (1988) **52**:479-487; [b]Tsallis C. Entropy in: *Encyclopedia of complexity and systems science*, R.A. Meyers (Ed.), Springer, New York (2009); [c]Tsallis C. *Introduction to nonextensive statistical mechanics; Approaching a Complex World*, Springer-Berlin (2009).

14. Greer J.R., de Hosson J.Th.M. Plasticity in small-sized metallic systems: Intrinsic versus extrinsic size effect, *Prog. Mat. Sci.* (2011) **56**:654-724.

15. [a]Aifantis K.E., Hackney S.A., Kumar V.R. (Eds.) *High Energy Density Lithium Batteries: Materials, Engineering, Applications*, Wiley-VCH (2010). [also translated in Chinese: China Machine Press, ISBN: 9787111371786, 2011]; [b]Ryu I., Choi J.W., Cui Y., Nix Y. Size-dependent fracture of Si nanowire battery anodes, *J. Mech. Phys. Solids* (2011) **59**:1717-1730; [c]Cui Z., Gao F., Qu J., Interface-reaction controlled diffusion in binary solids with applications to lithiation of silicon in lithium-ion batteries, *J. Mech. Phys. Solids* (2013) **61**:293-310; [d]Cheng Y.T., Verbrugge M.W., Desphande R., Understanding diffusion-induced stresses in lithium ion battery electrodes, in: *IUTAM Symposium on Surface Effects in the Mechanics of Nanomaterials and Heterostrucures*, Kocks A., Wang J. (Eds.), Springer, Dordrecht (2013) pp. 203-214.

16. [a]Walgraef D., Aifantis E.C. Dislocation patterning in fatigued metals as a result of dynamical instabilities, *J. Appl. Phys.* (1985) **58**:688-691; [b]Pontes J., Walgraef D., Aifantis E.C. On dislocation patterning: Multiple slip effects in the rate equation approach, Int. J. Plasticity (2006) **22**:1486-1505; [c]Spiliotis K.G., Russo L., Siettos C., Aifantis E.C. Analytical and numerical bifurcation analysis of dislocation pattern formation of the Walgraef-Aifantis model, *Int. J. Non-Linear Mech.* (2018) **102**:41-52.

17. [a]Hatzikirou H., Aifantis E.C., On the similarities between the W-A model for dislocations and the GoG model for cancer cells, Forthcoming (2019); [b]Hatzikirou H., Basanta D., Simon M., Schaller K., Deutsch A. 'Go or Grow': the key to the emergence of invasion in tumour progression?, *Math. Med.*

*Biol.* (2010) **29**:49–65; [c]Boettger K., Hatzikirou H., Voss-Böhme A., Ada Cavalcanti-Adam E., Herrero M.A., Deutsch A., An emerging Allee effect is critical for tumor initiation and persistence, *PLOS Comp. Biol.* (2015) **11**:E1004366.

18. Murray JD. *Mathematical Biology ; Vol I – An Introduction*, Springer, New York, 2002 ; *Vol II – Spatial Models and Biomedical Applications*, Springer, New York, (2003) [corrected Second Printing, 2004].

19. [a]Aifantis E.C., Hirth J.P. (Eds.) *The Mechanics of Dislocations*, ASM, Metals Park (1985); [b]Aifantis E.C., Walgraef D., Zbib H.M. (Eds.) *Material Instabilities*, Special Issue of  Res Mechanica, Vol. 23/ Nos. 2-3/ pp. 97-305, Elsevier Appl. Sci. Publ. (1988); [c]Estrin Y., Kubin L.P., Aifantis E.C., Introductory remarks to the  viewpoint set in propagative plastic instabilities, *Scripta Met. Mater.* (1993) **29**:1147-1150.

20. Aifantis E.C. Update on a class of gradient theories, *Mech. Mater.* (2003) **35**:259-280.

21. [a]Kubin L.P. In: *Plastic Deformation and Fracture of Materials*, H. Mughrabi (Ed.), Weinheim (FRG): VCH (1993) pp. 137-190; [b]Kubin L.P., Fressengeas C., Ananthakrishna G. *Dislocations in Solids*, Vol. 11, F.R.N. Nabarro (Ed.), Amsterdam, North-Holland (2002) pp. 101-192; [c]Ananthakrishna G. *Phys. Rep.* (2007) **440**:113-259; [d]Sauzay M., Kubin L.P. Scaling laws for dislocation microstructures in monotonic and cyclic deformation of fcc metals, *Prog. Mater. Sci.* (2011) **56**:725-784.

22. [a]A. Carpinteri (Ed.), *Size-Scale Effects in the Failure Mechanisms of Materials and Structures*, in: Proc. IUTAM Symp., Torino, Italy, 1994, E & FN SPON, London, 1996; [b]H.-B. Muhlhaus (Ed.), *Continuum Models for Materials with Microstructure*, Wiley, Chichester, 1995; [c]R. de Borst and E. van der Giessen (Eds.), *Material Instabilities in Solids*, Wiley, Chichester, 1998.

23. [a]Gutkin M.Yu., Aifantis E.C. Dislocations and disclinations in gradient elasticity, *Phys. Stat. Sol. B* (1999) **214**:245-284; [b]Lazar M., Maugin G.A., Aifantis E.C. Dislocations in second strain gradient elasticity, *Int. J. Sol. Struct.* (2006) **43**:1787-1817; [c]Askes H., Aifantis E.C. Gradient elasticity in statics and dynamics: an overview of formulations, length scale identification procedures, finite element implementations and new results, *Int. J. Solids Struct.* (2011) **48**:1962-1990; [d]Aifantis E.C., On non-singular GRADELA crack fields, *Theor. App. Mech. Lett.* (2014) **4**:051005.

24. [a]Aifantis E.C., Gittus J. (Eds.), *Phase Transformations*, Elsevier Appl. Sci. Publ., London-New York (1986); [b]Suresh S. *Fatigue of Materials*, Cambridge Univ. Press, Cambridge (1991); [c]Walgraef D. *Spatio-Temporal Pattern Formation*, Springer, New York (1997); [d]Gutkin M.Yu., Ovid'ko I.A. *Plastic Deformation in Nanocrystalline Materials*, Springer, Berlin (2004); [e]Ghoniem N., Walgraef D. *Instabilities and Self-Organization in Materials*, Vols. I & II, Oxford Sci. Publ., Oxford (2008); [f]Gurtin M.E., Fried E., Anand L. *The Mechanics and Thermodynamics of Continua*, Cambridge Univ. Press, New York (2010).

25. Po G., Lazar M., Seif D., Ghoniem N. Singularity-free dislocation dynamics with strain gradient elasticity, *J. Mech. Phys. Solids* (2014) **68**:161-178.

26. [a]Isaksson P., Dumont P.J.J., Rolland du Roscoat S. Crack growth in planar elastic fiber materials, *Int. J. Solids Struct.* (2012) **49**:1900-1907; [b]Isaksson P., Hägglund R., Crack-tip fields in gradient enhanced elasticity, *Engng Fract. Mech.* (2013) **97**:186-192.

27. Bagni C., Askes H., Aifantis E.C. Gradient-enriched finite element methodology for axisymmetric problems, *Acta Mech.* (2017) **228**:1423-1444.

28. Tsagrakis I., Aifantis E.C. Gradient elasticity effects on the two-phase lithiation of LiB anodes, in: *Generalized Models and Non-classical Approaches in Complex Materials 2, Adv. Struct. Mat.* **90**, H. Altenbach et al (Eds.), Springer (2018), pp. 221-235.

29. Spiliotis K.G., Russo L., Siettos C., Aifantis E.C. Latttice Boltzmann consideration in Dislocation Dynamics. An Equation Free approach, *Int. J. Numer. Met. In Eng.* (forthcoming).

30. [a]Konstantinidis A.A., Aifantis K.E., de Hosson J.Th.M. Capturing the stochastic mechanical behavior of micro and nanopillars, *Mater. Sci. Eng. A* (2014) **597**:89-94; [b]Konstantinidis A.A., Zhang X., Aifantis E.C. On the combined gradient-stochastic plasticity model: Application to Mo-micropillar compression, *AIP Conf. Proc.* (2015) **1646**:3-9.

31. [a]Zaiser M., Avlonitis M., Aifantis E.C. Stochastic and deterministic aspects of strain localization during cyclic plastic deformation, *Acta Mater.* (1998) **48**:4143-4151; [b]Avlonitis M., Zaiser M., Aifantis E.C. Some exactly solvable models for the statistical evolution of internal variables during plastic deformation, *Prob. Engng. Mech.* (2000) **15**:131.

32. [a]Chattopadhyay A.K., Aifantis E.C., Stochastically forced dislocation density distribution in plastic deformation, *Phys. Rev. E* (2016) **94**:022139; [b]Chattopadhyay A.K., Aifantis E.C., Double diffusivity model under stochastic forcing, *Phys. Rev. E* (2017) **95**:052134.

33. [a]Zaiser M., Aifantis E.C. Avalanches and Slip Patterning in Plastic Deformation, *J. Mech. Beh.Mat.* (2003) **14**:255-270; [b]Zaiser M., Aifantis E.C. Randomness and slip avalanches in gradient plasticity, *Int. J. Plasticity* (2006) **22**:1432-1455.

34. Li H., Ngan A.H.W., Wang M.G. Continuous strain bursts in crystalline and amorphous metals during plastic deformation by nanoindentation, *J. Mater. Res.* (2005) **20**:3072-3081.

35. [a]Iliopoulos A.C., Nikolaidis N.S., Aifantis E.C. Analysis of serrations and shear bands fractality in UFGs, *J. Mech. Behav. Mater.* (2015) **24**:1-9; [b]Iliopoulos A.C., Aifantis E.C. Tsallis q-triplet, intermittent turbulence and Portevin–Le Chatelier effect, *Physica A* (2018) **498**:17-32.

36. [a]Kawazoe H., Yoshida M., Basinski Z.S., Niewczas M. Dislocation microstructures in fine-grained Cu polycrystals fatigued at low amplitude, *Scripta Mater.* (1999) **40**:639-644; [b]Wang D., Volkert C.A., Kraft O. Effect of length scale on fatigue life and damage formation in thin Cu films, *Mat. Sci. Eng. A* (2008) **493**:267-273.

37. Unger D.J., Gerberich W.W., Aifantis E.C. Further remarks on the implications of steady state stress assisted diffusion on environmental cracking, *Scripta Metall.* (1982) **16**:1059-1064.

38. Silber G., Trostel R., Alizadeh M., Benderoth G. A continuum mechanical gradient theory with applications to fluid mechanics, *J. Phys. IV France* (1998) **8**:Pr8/365-373.

39. Fried E., Gurtin M.E. Tractions, Balances, and Boundary Conditions for Nonsimple Materials with Application to Liquid Flow at Small-Length Scales, *Arch. Rat. Mech. Anal.* (2006) **182**:513–554.

40. Adams, J.M., Fielding, S.M., Olmsted, P.D. The interplay between boundary conditions and flow geometries in shear banding: Hysteresis, band configurations, and surface transitions. *J. Nonnewton. Fluid Mech.* (2008) **151**:101-118.

41. Cates, M.E., Fielding, S.M. Rheology of giant micelles. *Adv. Phys.* (2006) **55**:799–879.

42. Dhont J.K.G., Briels W.J.: Gradient and vorticity banding. *Rheol. Acta.* (2008) **47**:257–281.

43. Ru C.Q., Aifantis E.C. A simple approach to solve boundary-value problems in gradient elasticity. *Acta Mech.* (1993) **101**:59–68.

44. Giusteri G.G., Fried E. Slender-body theory for viscous flow via dimensional reduction and hyperviscous regularization. *Meccanica.* (2014) **49**:2153–2167.

45. Vardoulakis I., Aifantis E.C. Gradient dependent dilatancy and its implications in shear banding and liquefaction, Ingenieur-Archiv (1989) **59**:197-208.

46. Vardoulakis I., Muhlhaus H.B., Aifantis E.C., Continuum models for localized deformations in pressure sensitive materials, in: Computer Methods and Advances in Geomechanics, G. Beer et al (Eds.), Balkema Publ., Rotterdam (1991), pp. 441-448.

47. Vardoulakis I., Aifantis E.C. A gradient flow theory of plasticity for granular materials, *Acta Mech.* (1991) **87**:197-217.

48. Vardoulakis I., Aifantis E.C. On the role of microstructure in the behavior of soils: Effects of higher order gradients and internal inertia, Mech. Mat. (1994) **18**:151-158.

49. Oka F., Yashima A., Sawada K., Aifantis E.C. Instability of gradient-dependent elastoviscoplastic model for clay and strain localization, *Comp. Meth. Appl. Mech. Engng.* (2000) **183**:67-86.

50. di Prisco C., Imposimato S., Aifantis E.C. A visco-plastic constitutive model for granular soils modified according to non-local and gradient approaches, *Int. J. Num. Anal. Meth. Geomech.* (2002) **26**:121-138.

51. Fyffe B., Schwerdtfeger J., Blackford J.R., Zaiser M., Konstantinidis A., Aifantis E.C. Fracture toughness of snow: The influence of layered microstructure, *J. Mech. Beh. Mat.* (2007) **18**:195-215.

52. Konstantinidis A., Cornetti P., Pugno N., Aifantis E.C. Application of gradient theory and quantized fracture mechanics in snow avalanches, *J. Mech. Behav. Mat.* (2009) **19**:39-48.

53. Haoxiang C., Qi C., Peng L., Kairui L., Aifantis E.C., Modeling the zonal disintegration of rocks near deep level tunnels by gradient internal variable continuous phase transition theory, *J. Mech. Behav. Mat.* (2015) **24**:161-171.

54. Qi C., Wei X., Hongsen W., Aifantis E.C. On temporal-structural dynamic failure criteria for rocks, *J. Mech. Behav. Mat.* (2015) **24**:173-181.

55. Efremidis G., Avlonitis M., Konstantinidis A., Aifantis E.C. A statistical study of precursor activity in earthquake-induced landslides, *Comput. Geotechn.* (2017) **81**:137-142.

56. Chen H., Qi C., Efremidis G., Dorogov M., Aifantis E.C. Gradient elasticity and size effect for the borehole problem, *Acta Mech.* (2018) **229**:3305-3318.

57. Ord A., Hobbs B.E. Fracture pattern formation in frictional, cohesive, granular material, *Philos. Trans. R. Soc. A Math. Phys. Eng. Sci.* (2010) **368**:95–118.

58. Yue Y.M., Xu K.Y., Aifantis E.C. Microscale size effects on the electromechanical coupling in piezoelectric material for anti-plane problem, *Smart Mater. Struct.* (2014) **23**:125043/1-11.

59. Yue Y.M., Xu K.Y., Chen T., Aifantis E.C. Size effects on magnetoelectric response of multiferroic composite with inhomogeneities, *Physica B* (2015) **478**:36-42.

60. Yue Y.M., Xu K.Y., Aifantis E.C. Strain gradient and electric field gradient effects in piezoelectric cantilever beams, *J. Mech. Behav. Mat.* (2015) **24**:121-127.

61. Tarasov V.E., Trujillo J.J. Fractional power-law spatial dispersion in electrodynamics. *Annals of Physics* (2013) **334**:1–23.

62. Truesdell C., Toupin R. The Classical Field Theories. In: Flügge S. (eds) Principles of Classical Mechanics and Field Theory/Prinzipien der Klassischen Mechanik und Feldtheorie. Encyclopedia of Physics/ Handbuch der Physik, vol 2/3/1, Springer, Berlin, Heidelberg (1960), pp 226-858.

63. Zimmerman J.A., Webb E.B., Hoyt J.J., Jones R.E., Klein P.A., Bammann D.J. Calculation of stress in atomistic simulation, *Modelling Simul. Mater. Sci. Eng.* (2004) **12**:S319-S332.

64. Maranganti R., Sharma P. Revisiting quantum notions of stress, *Proc. Royal Soc. A* (2010) **466**:2097–2116.

65. Davies H. *The physics of Low-Dimensional Semiconductors*, Cambridge University Press, Cambridge, (2000).

66. Zhang X., Gharbi M., Sharma P., Johnson H.T. Quantum field induced strains in nanostructures and prospects for optical actuation, *Int. J. Solids Struct.* (2009) **46**:3810-3824.

67. Vayenas C.G., Souentie S. *Gravity, Special Relativity, and the Strong Force*. Springer US, Boston, MA (2012).

68. Vayenas C.G., Souentie S., Fokas A. A Bohr-type model of a composite particle using gravity as the attractive force. *Phys. A Stat. Mech. its Appl.* (2014) **405**:360–379.

69. London F., Zur Theorie und Systematik der Molekularkräfte, *Z. Physik* (1930) **63**:245–279.

70. London F. The general theory of molecular forces, *Trans. Faraday Soc.* (1937) **33**:8–26.

71. Jones J.E. On the Determination of Molecular Fields — I. From the Variation of the Viscosity of a Gas with Temperature, *Phil. Trans. A* (1924) **106**:441–462.

72. Israelachvili J.N. Intermolecular and Surface Forces 3ed, Academic Press (2011).

73 Parson J.M., Siska P.E., Lee Y.T. Intermolecular potentials from crossed-beam differential elastic scattering measurements. IV. Ar+Ar, *J. Chem. Phys.* (1972) **56**:1511–1516.

74 Stillinger F.H., Weber T.A. Computer simulation of local order in condensed phases of silicon, *Phys. Rev. B* (1985) **31**:5262–5271.

75. M. Lazar, G.A. Maugin, Aifantis E.C. On the theory of nonlocal elasticity of bi- Helmholtz type and some applications, *Int. J. Solids. Struct.* (2006) **43**:1404-1421.

76. Kioseoglou J., Dimitrakopulos G.P., Komninou Ph., Karakostas T., Aifantis E.C. Dislocation core investigation by geometric phase analysis and the dislocation density tensor, *J. Phys. D: Appl. Phys.* (2008) **41**:035408/1-8.

77. Aifantis E.C. Non-singular dislocation fields, *IOP Conf. Series: Mat. Sci. Eng.* (2009) **3**:0712026.

78. [a]Tarasov V.E., Aifantis E.C. Toward fractional gradient elasticity, *J. Mech. Beh. Mat.* (2014) **23**:41-46; [b]Tarasov V.E., Aifantis E.C. Non-standard extensions of gradient elasticity: Fractional non-locality, memory and fractality, *Commun. Nonlinear Sci. Numer. Simulat.* (2015) **22**:197-227; [c] Aifantis E.C. Fractional generalizations of gradient mechanics in: Handbook of Fractional Calculus with Applications Volume 4, V.E. Tarasov (Ed), De Gruyter, Berlin (2019); [d]Tarasov V.E., Aifantis E.C. On fractional and fractal formulations of gradient linear and nonlinear elasticity, *Acta Mech.* (2019) **230**:2043–2070; [e]Parisis K., Konstantopoulos I., Aifantis E.C. Nonsingular solutions of GradEla models for dislocations: An extension to fractional GradEla, *J. Micromech. Mol. Phys.* (2018) **3**:1840013/1-13.

79. [a]Samko S., Kilbas A., Marichev O. *Integrals and Derivatives of Fractional Order and Applications*, New York: Gordon and Breach (1993); [b]Kilbas A., Srivastava M., Trujillo J. *Theory and Applications of Fractional Differential Equations*, Elsevier (2006); [c]Mathai A., Saxena R.K., Haubold H.J. *The H-Function: Theory and Applications*, Springer-Verlag New York (2010).

80. Eringen, A.C. *Microcontinuum Field Theories I: Foundations and Solids*, Springer, New York (1999).

81. [a]Fleck N.A., Hutchinson J.W. Strain gradient plasticity. *Adv. Appl. Mech.* (1997) **33**:295–361; [b]Fleck N.A., Hutchinson J.W. A reformulation of strain gradient plasticity. *J. Mech. Phys. Solids* (2001) **49**:2245–2271.

82. Gurtin M.E., Anand L. Thermodynamics applied to gradient theories involving the accumulated plastic strain: the theories of Aifantis and Fleck and Hutchinson and their generalization. *J. Mech. Phys. Solids* (2009) **57**:405–421.

83. [a]Gao H., Huang Y., Nix W.D., Hutchinson J.W. Mechanism-based strain gradient plasticity - I. Theory. *J. Mech. Phys. Solids* (1999) **47**:1239–1263; [b]Nix W.D., Gao H.J. Indentation size effects in crystalline materials: a law for strain gradient plasticity. *J. Mech. Phys. Solids* (1998) **46**:411–425.

84. [a]de Borst R., Muhlhaus H.B. Gradient-dependent plasticity - formulation and algorithmic aspects. *Int. J. Numer. Methods Eng.* (1992) **35**:521–539; [b]de Borst R., Pamin J., Sluys L.J. Computational Issues in Gradient Plasticity. In: H-B. Mühlhaus (Ed.), *Continuum models for materials with microstructure*, Wiley (1995). pp. 159–200.

85. [a]Geers M.G.D., Peerlings R.H.J., Brekelmans W.A.M., de Borst R. Phenomenological nonlocal approaches based on implicit gradient-enhanced damage. *Acta Mech.* (2000) **144**:1–15; [b]Peerlings R.H.J., Poh L.H., Geers M.G.D. An implicit gradient plasticity-damage theory for predicting size effects in hardening and softening. *Eng. Fract. Mech.* (2012) **95**:2–12.

86. [a]Willis J.R. Some forms and properties of models of strain-gradient plasticity. *J. Mech. Phys. Solids* (2019) **123**:348–356; [b]Aifantis K.E., Willis J.R. The role of interfaces in enhancing the yield strength of composites and polycrystals. *J. Mech. Phys. Solids* (2005) **53**:1047–1070.

87. [a]Polizzotto C. Unified thermodynamic framework-for nonlocal/gradient continuum theories. *Eur. J. Mech. A Solid.* (2003) **22**:651–668; [b]Polizzotto C., Interfacial energy effects within the framework of strain gradient plasticity. *Int. J. Solids Struct.* (2009) **46**:1685–1694.

88. Voyiadjis G.Z., Song Y. Strain gradient continuum plasticity theories: Theoretical, numerical and experimental investigations. *Int. J. Plast.* (2019) **121**:21–75.

89. [a]Goddard J.D., On linear non-local thermo-viscoelastic waves in fluids. *Mat. Mech. Compl. Sys.* (2018) **6**:321–338; [b]Goddard J.D. On the stability of the μ(I) rheology for granular flow. *J. Fluid Mech.* (2017) **833**:302–331.

90. [a]Kamrin, K., Koval, G. Nonlocal constitutive relation for steady granular flow. *Phys. Rev. Lett.* (2012) **108**:178301; [b]Henann D.L., Kamrin K. A predictive, size-dependent continuum model for dense granular flows. *Proc. Natl. Acad. Sci. U. S. A.* (2013) **110**:6730-6735.

91. Forterre Y., Pouliquen O. Flows of dense granular media. *Annu. Rev. Fluid Mech.* (2008) **40**:1–24.

92. Fenistein D., van Hecke M. Wide shear zones in granular bulk flow. *Nature.* (2003) **425**:256–256.

93. Dijksman J.A., Wortel G.H., van Dellen L.T.H., Dauchot O., van Hecke M. Jamming, Yielding, and Rheology of Weakly Vibrated Granular Media. *Phys. Rev. Lett.* (2011) **107**:108303.

94. Bocquet L., Colin A., Ajdari A. Kinetic theory of plastic flow in soft gassy materials. *Phys. Rev. Lett.* (2009) **103**:036001.

95. [a]Fischbach E., Sudarsky D., Szafer A., Talmadge C., Aronson S.H. Reanalysis of the Eötvös experiment. *Phys. Rev. Lett.* (1986) **56**:3–6; [b]Fischbach E., The fifth force: A personal history. *Eur. Phys. J. H.* (2015) **40**:385–467.